%% file: main_arxiv.tex
\documentclass[10pt,conference]{IEEEtran}

\usepackage{amsmath}
\usepackage{amssymb}
\usepackage{graphicx}
\usepackage{fink}
\usepackage[table]{xcolor}
\usepackage[font=footnotesize]{caption}
\usepackage{stmaryrd}
\usepackage{multirow}
\usepackage{mathpartir}
\usepackage{amsthm}
\usepackage{enumerate}
\usepackage{subfig}
\usepackage{bm}
\usepackage{float}
\usepackage{hyperref}
\usepackage{algorithm,algorithmic}

\newcommand{\dist}[1]{\text{\em Dist}(#1)}

\renewcommand{\L}{\mathcal{L}}

\newcommand{\dirac}[1]{\delta_{#1}}
\newcommand{\ie}{{\em i.e. }}
\newcommand{\viz}{{\em viz. }}

\newcommand{\preorder}{\preceq}
\newcommand{\dpreorder}{\sqsubseteq}
\newcommand{\trans}[1]{\overset{#1}{\rightarrow}}

\newcommand{\supp}[1]{\text{{\em Supp}}(#1)}

\newcommand{\bigor}{\bigvee}
\newcommand{\bigand}{\bigwedge}
\newcommand{\pcex}{\mathcal{P}}
\newcommand{\ncex}{\mathcal{N}}

\newcommand{\lift}[1]{\text{{\em lift}}(#1)}

\newcommand{\rat}{\mathbb{Q}}
\newcommand{\product}{\otimes}

\newcommand{\parent}[1]{\text{{\em par}}(#1)}
\newcommand{\act}[1]{\text{{\em act}}(#1)}
\newcommand{\poly}[1]{\text{{\em poly}}(#1)}

\newtheorem{definition}{{\bf Definition}}
\newtheorem{lemma}{{\bf Lemma}}
\newtheorem{theorem}{{\bf Theorem}}
\newtheorem{corollary}{{\bf Corollary}}
\newtheorem{condition}{{\bf Condition}}

\IEEEoverridecommandlockouts

\begin{document}

\title{Learning Probabilistic Systems from Tree Samples
\thanks{
This research was sponsored by DARPA META II, GSRC, NSF, 
SRC, GM, ONR under contracts FA8650-10C-7079,
1041377 (Princeton University), CNS0926181/CNS0931985, 2005TJ1366, 
GMCMUCRLNV301, N000141010188, respectively, and the CMU-Portugal Program.
{\em Original Publication:} A. Komuravelli, C.S. P\u{a}s\u{a}reanu and E.M.
Clarke. Learning Probabilistic Systems from Tree Samples. In proceedings of LICS, pp. 441-450,
\copyright\ 2012 IEEE, available at \url{http://dx.doi.org/10.1109/LICS.2012.54}.
}
}

\author{\IEEEauthorblockN{Anvesh Komuravelli}
\IEEEauthorblockA{Computer Science Department \\ Carnegie Mellon University \\
Pittsburgh, PA, USA}
\and
\IEEEauthorblockN{Corina S. P\u{a}s\u{a}reanu}
\IEEEauthorblockA{Carnegie Mellon Silicon Valley \\ NASA Ames Research Center\\ Moffett Field,
CA, USA}
\and
\IEEEauthorblockN{Edmund M. Clarke}
\IEEEauthorblockA{Computer Science Department \\ Carnegie Mellon University \\
Pittsburgh, PA, USA}}



\maketitle

\input{abstract}
\input{introduction}

\input{preliminaries}
\input{learning_consistent}
\input{learning_lpts}

\input{learning_assumptions}
\input{conclusion}
\input{ack_arxiv}




\input{main_arxiv.bbl}

\newpage
\appendix
\input{appendix}

\end{document}

%% file: abstract.tex
\begin{abstract}
We consider the problem of learning a non-deterministic probabilistic
system consistent with a given finite set of positive and negative
tree samples.  Consistency is defined with respect to strong
simulation conformance. We propose learning algorithms that use
traditional and a new {\em stochastic} state-space partitioning, the
latter resulting in the minimum number of states.  We then use them to
solve the problem of {\em active learning}, that uses a knowledgeable
teacher to generate samples as counterexamples to simulation 
equivalence queries. We show that the problem is undecidable
in general, but that it becomes decidable under a suitable condition
on the teacher which comes naturally from the way samples are
generated from failed simulation checks.  The latter problem is shown
to be undecidable if we impose an additional condition on the learner
to always conjecture a {\em minimum state} hypothesis. We therefore
propose a semi-algorithm using stochastic partitions. Finally, we
apply the proposed (semi-) algorithms 
to infer intermediate assumptions in an automated assume-guarantee 
verification framework for probabilistic systems.
\end{abstract}

\begin{IEEEkeywords}
probability, transition, system, simulation, conformance, active learning,
tree, partition, assume-guarantee
\end{IEEEkeywords}

%% file: introduction.tex
\section{Introduction}

We study the problem of learning an unknown non-deterministic {\em
  Labeled Probabilistic Transition System} (LPTS) from tree
samples. The motivation for this work was to investigate learning
techniques for automating assume-guarantee style~\cite{Pnueli_lmcs85} compositional
verification of strong simulation conformance~\cite{SL_nordic95}
between LPTSes. Strong simulation for LPTSes is decidable in
polynomial time~\cite{BEM_jcss00} and yields {\em stochastic
  tree} counterexamples when it fails~\cite{KPC_cav12}. Stochastic trees
are {\em tree-shaped} LPTSes (see Section \ref{sec:prelims}) with
probabilities appearing on the transitions.

Compositional verification~\cite{CLM_lics89} is a promising approach for alleviating the
state explosion problem in model checking~\cite{CGP_book}. Learning from
trace~\cite{Angluin_infcomp87,OM_spire98} and tree~\cite{CCS+_cav05}
counterexamples has been successfully applied before for automating
the approach in a non-probabilistic setting, for checking trace
inclusion~\cite{PGB+_fmsd08,CFC+_tacas09} and simulation conformance~\cite{CCS+_cav05},
respectively. The most closely related work~\cite{CCS+_cav05}
reduces simulation conformance to {\em tree language} inclusion and
uses learning for deterministic tree automata to automatically generate the assumptions
used in compositional reasoning.
In the probabilistic setting,  existing literature has dealt with
learning from samples consisting of trees with information regarding
the probability of acceptance~\cite{COC_ml01}, but learning from
stochastic trees has not been considered before. Moreover, there is no existing
probabilistic variant of a tree automaton to recognize stochastic tree
languages.
This motivated us to consider learning an LPTS directly, without 
working with tree languages or tree automata.

We consider first the problem of learning a non-deterministic LPTS
that is {\em consistent} with respect to a set of positive and negative
stochastic tree samples, where consistency is defined in terms of 
strong simulation conformance. For the purpose of verification, we
want the learnt models to be {\em minimal} or at least to have a good
upper bound on their size.  We describe two algorithms, each
using a different way of partitioning the state-space of the positive
samples. One algorithm uses traditional state-space partitioning
(Section \ref{sec:partitions}) resulting in the least number of
partitions, while the other uses a new {\em stochastic} partitioning
(Section \ref{sec:stochastic_partitions}) resulting in the least
number of states.

We then apply the above algorithms to solve the problem of
learning an unknown target in Section \ref{sec:learning_lpts}. This is
done in the framework of {\em active learning} with the help of a
knowledgeable teacher.  Typically active learning algorithms assume a
teacher that answers two types of queries - {\em membership} (of a
sample in the unknown target) and {\em equivalence} (between the
conjectured model and the unknown target)~\cite{Angluin_infcomp87}.
However we observe that membership queries are not straightforward to
create in our case as the learner would need to guess the transition
probabilities, along with the tree-structure. Therefore, we only
assume the teacher can answer equivalence queries -- the teacher checks
simulation equivalence (two-way simulation conformance) between a
conjectured LPTS and the target LPTS and returns positive or negative
stochastic trees when the check fails.

We show that active learning for LPTSes is undecidable in general. We
then propose a learning algorithm that works under an assumption on
the teacher which comes naturally from the way the tree
counterexamples are generated from failed simulation checks.  As we
are interested in learning an LPTS of the least number of states, we
also consider imposing a restriction on the learner to always
conjecture a {\em minimum state} hypothesis. Learning with this restriction
also turns out to be undecidable and we propose a semi-algorithm using
stochastic partitions.

LPTSes are related to {\em probabilistic automata}
(PA)~\cite{Rabin_infctrl63}. Algorithms to learn PAs have only been proposed in restricted settings of stronger
assumptions on a teacher~\cite{Tzeng_ml92} or approximate
learning~\cite{HO_icgi04,MCJ+_qest11}.
Algorithms to learn a {\em multiplicity} automaton,
which generalizes a PA by replacing the probabilities with arbitrary
rationals, have also been proposed~\cite{BBB+_jacm00}. 
Adapting these to solve verification problems involving probabilistic
transition systems is difficult and results in non-terminating
algorithms~\cite{FHK+_atva11}. On the other hand, we show in Section
\ref{sec:ag} that one can readily apply the algorithms we propose to
infer intermediate assumptions in an automated assume-guarantee style
framework for the verification of strong simulation conformance
between LPTSes. This yields the first complete and fully automated learning
framework for compositional verification of probabilistic systems.
Moreover, one can extend this framework to check logical
properties, such as the fragment {\em weakly safe PCTL}~\cite{CV_tocl10}, which are
preserved by the conformance and also have tree counterexamples.

\noindent{\bf Other Related Work.}
Learning for automating compositional reasoning of probabilistic
systems has been proposed before~\cite{FKP_fase11} in the context of
checking probabilistic reachability properties, which are refuted by
sets of trace counterexamples.  The approach uses a variant of L*~\cite{Angluin_infcomp87},
a learning algorithm for DFAs, to
automatically learn deterministic assumptions, following previous work
in the non-probabilistic setting~\cite{PGB+_fmsd08}. The approach uses
a sound but incomplete rule, and therefore, it is not guaranteed to
terminate (completeness is necessary for termination).  A
complete rule for such properties restricted to systems without
non-determinism has been considered recently~\cite{FHK+_atva11}. It
uses learning with {\em probabilistic} trace inclusion as the
conformance relation which is undecidable. Also, the learning algorithm is not
guaranteed to terminate. In contrast, we use simulation conformance
which is decidable in polynomial time and leads to a sound
and complete rule (Section \ref{sec:ag}). We are also able to
guarantee termination for the algorithm proposed in Section
\ref{sec:ag} when using classical partitions to infer a
consistent LPTS.


Our work draws inspiration from a previous work~\cite{GMF_fmsd08} that
automates assumption generation by using an algorithm for learning the
{\em minimal separating automaton} from positive and negative trace
counterexamples. The counterexamples are provided via model checking
in an assume-guarantee framework.
Similar to our work, they use a {\em
  partitioning approach}, where the goal is to find a {\em folding} of
the counterexamples into the learnt model.  A different approach has
been proposed to find the separating automaton based on L* which makes
use of membership queries, in addition to equivalence queries~\cite{CFC+_tacas09}. 
All these works were done in the context of
non-probabilistic reasoning under trace semantics and thus, are 
different from our setting.

Learning a minimum-state automaton from positive and negative samples
is a well studied problem~\cite{AS_compsurv83,OG_asspr92,GO_report93}
that is known to be hard~\cite{Gold_infcomp78}.  Algorithms have also
been proposed for samples with stochastic information, \ie the
probability of acceptance of a trace or a tree~\cite{CO_rairo99,COC_ml01},
learning stochastic finite (tree)
automata.  
As also previously said, we cannot
immediately borrow existing results from the above automata-theoretic
approaches.

%% file: preliminaries.tex
\section{Preliminaries}
\label{sec:prelims}

\noindent{\bf Labeled Probabilistic Transition Systems.}
Let $S$ be a non-empty set. $\dist{S}$ is defined
to be the set of discrete probability distributions over $S$. 
We assume that all the probabilities specified explicitly in a distribution are
rationals in $[0,1]$; there is no unique
representation for all real numbers on a computer and floating-point numbers are
essentially rationals.
For $s \in S$, $\dirac{s}$ is the Dirac distribution on $s$, \ie
$\dirac{s}(s) = 1$ and $\dirac{s}(t) = 0$ for all $t \neq s$. For
$\mu \in \dist{S}$, the {\em support} of $\mu$,
denoted $\supp{\mu}$, is defined to be the set $\{s \in S |
\mu(s) > 0\}$ and for $X \subseteq S$, $\mu(X)$ stands for
$\sum_{s \in X} \mu(s)$.
The models we consider, defined below, have both probabilistic and
non-deterministic behavior. Thus, there can be a non-deterministic choice
between two probability distributions, even for the same action. Such modeling
is typically used for underspecification.
Moreover, the theory described does not become any simpler by disallowing non-deterministic choice for a given
action (see the discussion on counterexamples at the end of this section).

\begin{definition}[LPTS]
\label{def:lpts}
A {\em Labeled Probabilistic Transition System} $($LPTS$)$ is a tuple $\langle
S, s^0, \alpha, \tau \rangle$ where $S$ is a set of states, $s^0
\in S$ is a distinguished start state, $\alpha$ is a set of actions and $\tau
\subseteq S \times \alpha \times \dist{S}$ is a probabilistic transition
relation. For $s \in S$, $a \in \alpha$ and $\mu \in \dist{S}$, we denote
$(s,a,\mu) \in \tau$ by $s \trans{a} \mu$ and say that $s$ has a
{\em transition} on $a$ to $\mu$.

An LPTS is called {\em reactive} if $\tau$ is a partial function from $S \times
\alpha$ to $\dist{S}$ $(${\em \ie}at most one transition on a given action from a given
state$)$.
\end{definition}

\begin{figure}
\centering
\includegraphics[scale=1.6]{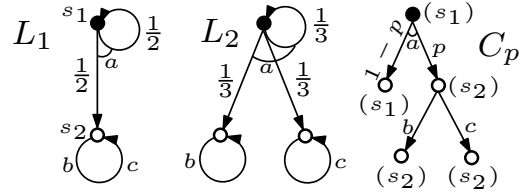}
\caption{Three reactive LPTSes. $p \in (0,1)$ for $C_p$.}
\label{fig:cex_eg}
\vspace{-0.2in}
\end{figure}

Throughout this paper, we use filled circles to denote start states in the pictorial representations of
LTPSes. For example, Figure \ref{fig:cex_eg} shows three LPTSes. For $\mu =
\{(s_1,\frac{1}{2}),(s_2,\frac{1}{2})\}$, $L_1$ has the transition $s_1
\trans{a} \mu$. All the LPTSes in the figure are {\em reactive} as no state has more than one transition on a given
action.
In the literature, an LPTS is also called a {\em
simple probabilistic automaton}~\cite{SL_nordic95}. Similarly, a
reactive LPTS is also called a (Labeled) {\em Markov Decision Process}.
Also, note that an LPTS with all the distributions
restricted to Dirac distributions is the classical (non-probabilistic)
{\em Labeled Transition System} (LTS); thus a {\em reactive} LTS
corresponds to the standard notion of a {\em deterministic}
LTS. We only consider finite state, finite
alphabet and finitely branching (\ie finitely many transitions from
any state) LPTSes. We use $\langle S_i, s^0_i, \alpha_i, \tau_i
\rangle$ for an LPTS $L_i$ and $\langle S_L, s^0_L, \alpha_L, \tau_L \rangle$
for an LPTS $L$.

We are also interested in LPTSes with a tree structure, \ie the start state is
not in the support of any distribution and every other state is in the support
of exactly one distribution. We call such LPTSes {\em stochastic trees} or
simply {\em trees}. For example, $C_p$, $p \in (0,1)$, in Figure
\ref{fig:cex_eg} is a tree.

%
%

\noindent{\bf Strong Simulation.}  In the non-probabilistic case, for two
labeled transition systems (LTSes), a pair of states belonging to a
strong simulation relation depends on whether certain other pairs of
successor states also belong to the relation~\cite{Milner_tr71}. For LPTSes, 
one has successor {\em distributions} instead of successor
states; a pair of states belonging to a strong simulation relation
$R$ should now depend on whether certain other pairs in the {\em
  supports} of these successor distributions also belong to $R$.
We thus need a binary relation between distributions,
$\dpreorder_R$, which depends on the relation $R$ between states.
Intuitively, two distributions can be related if we can pair the states in their
support sets, the pairs contained in $R$, {\em matching all} the probabilities under the
distributions.

\begin{figure}
\centering
\includegraphics[scale=1.6]{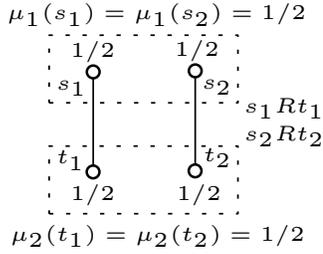}
\caption{A simple example where matching probabilities (solid edges) directly proves $\mu_1
\dpreorder_R \mu_2$.}
\label{fig:dist_rel_simple}
\end{figure}

\begin{figure}
\centering
\includegraphics[scale=1.6]{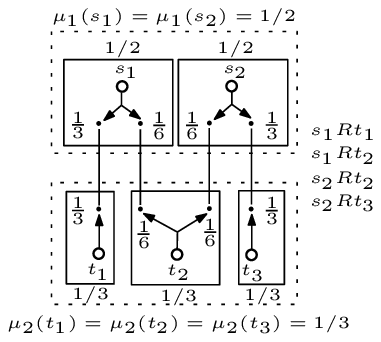}
\caption{An example where probabilities are split (arrows) before matching
(solid edges) to prove $\mu_1 \dpreorder_R \mu_2$.}
\label{fig:dist_rel_general}
\vspace{-0.2in}
\end{figure}

Consider an example with $s R t$ and the transitions $s \trans{a}
\mu_1$ and $t \trans{a} \mu_2$ with $\mu_1$ and $\mu_2$ as in Figure
\ref{fig:dist_rel_simple}. In this case, one easy way to match the probabilities is to pair $s_1$ with $t_1$ and
$s_2$ with $t_2$. This is sufficient if $s_1 R t_1$ and $s_2 R t_2$ also hold,
in which case, we say that $\mu_1 \dpreorder_R \mu_2$.
However, such a direct matching may not be possible in general. As shown in
Figure \ref{fig:dist_rel_general}, we need a more general
notion of matching the probabilities. One can
achieve that by {\em splitting} the probabilities under the distributions in such a way
that one can then directly match the probabilities as in Figure
\ref{fig:dist_rel_simple}. Now, if $s_1 R t_1$, $s_1 R
t_2$, $s_2 R t_2$ and $s_2 R t_3$ also hold, we say that $\mu_1 \dpreorder_R \mu_2$.
Note that there can more than one possible splitting.

This is the central idea behind the following definition where the splitting is achieved by a {\em
weight function}. For the rest of the section, let $L_1$ and $L_2$ be two
LPTSes, $\mu_1 \in \dist{S_1}$, $\mu_2 \in \dist{S_2}$ and $R \subseteq S_1
\times S_2$.

\begin{definition}[\cite{SL_nordic95}]
\label{def:weight_function_based}
$\mu_1 \dpreorder_R \mu_2$ iff there is a {\em weight} function $w :
S_1 \times S_2 \rightarrow \rat \cap [0,1]$ such that
\begin{enumerate}
  \item $\mu_1(s_1) = \sum_{s_2 \in S_2} w(s_1, s_2)$ for all $s_1 \in S_1$,
  \item $\mu_2(s_2) = \sum_{s_1 \in S_1} w(s_1, s_2)$ for all $s_2 \in S_2$,
  \item $w(s_1, s_2) > 0$ implies $s_1 R s_2$ for all $s_1 \in S_1$, $s_2
\in S_2$.
  \end{enumerate}
\end{definition}

$\mu_1 \dpreorder_R \mu_2$ can be checked by computing the maxflow in
an appropriate network and checking if it equals $1.0$~\cite{BEM_jcss00}.
If $\mu_1 \dpreorder_R \mu_2$ holds, $w$ in the above definition is one such
maxflow function.
As explained above, $\mu_1 \dpreorder_R \mu_2$ can be understood as {\em
matching} all the probabilities (after splitting appropriately) under $\mu_1$ and $\mu_2$. Considering
$\supp{\mu_1}$ and $\supp{\mu_2}$ as two partite sets, this is the weighted analog
of saturating a partite set in bipartite matching, giving us the following
analog of the well-known Hall's Theorem for saturating
$\supp{\mu_1}$.

\begin{lemma}[\cite{Zhang_thesis08}]
\label{lem:image_based}
$\mu_1 \dpreorder_R \mu_2$ iff for every $S \subseteq \supp{\mu_1}$, $\mu_1(S)
\leq \mu_2(R(S))$.
\end{lemma}

It follows that when $\mu_1 \not\dpreorder_R \mu_2$, there exists a witness $S
\subseteq \supp{\mu_1}$ such
that $\mu_1(S) > \mu_2(R(S))$. For example, if $R(s_2) = \emptyset$ in Figure
\ref{fig:dist_rel_simple}, its probability $\frac{1}{2}$ under $\mu_1$ cannot be
matched and $S = \{s_2\}$ is a witness subset.

\begin{definition}[Strong Simulation~\cite{SL_nordic95}]
\label{def:strong_simulation}
$R$ is a {\em strong simulation} iff for every $s_1 R s_2$ and $s_1 \trans{a}
\mu^a_1$ there is a $\mu^a_2$ with $s_2 \trans{a} \mu^a_2$ and
$\mu^a_1 \dpreorder_R \mu^a_2$.

For $s_1 \in S_1$ and $s_2 \in S_2$, $s_2$ strongly simulates $s_1$, denoted $s_1
\preorder s_2$, iff there is a strong simulation $T$ such that $s_1 T s_2$.  
$L_2$ strongly simulates $L_1$, also denoted $L_1 \preorder L_2$, iff $s^0_1
\preorder s^0_2$. For the latter, alternatively, we say that {\em simulation conformance} holds
between $L_1$ and $L_2$.
\end{definition}

\begin{definition}[Strong Simulation Equivalence]
\label{def:simeq}
The {\em strong simulation equivalence}, denoted $\simeq$, is defined as the kernel of strong
simulation, \ie $\simeq = \preorder \cap \succeq$.
\end{definition}

Definition \ref{def:strong_simulation} generalizes the one in the
non-probabilistic setting~\cite{Milner_tr71} and has the following immediate
consequence.

\begin{lemma}
\label{lem:css}
$\preorder \subseteq S_1 \times S_2$ is the coarsest strong simulation, \ie $\preorder$ is a strong
simulation and contains every strong simulation.
\end{lemma}

Simulation conformance is decidable in polynomial time~\cite{BEM_jcss00} and can be checked with
a greatest fixed point algorithm that computes the coarsest simulation
between $L_1$ and $L_2$. The algorithm uses a relation variable $R$
initialized to $S_1 \times S_2$ and it checks the condition in
Definition~\ref{def:strong_simulation} for every pair in $R$, iteratively,
removing any violating pairs from $R$. The
algorithm terminates when a fixed point is reached showing $L_1
\preorder L_2$ or when the pair of start states is removed showing
$L_1 \not\preorder L_2$.
Several optimizations exist~\cite{Zhang_thesis08} but we do not consider them
here, for simplicity.

\begin{lemma}[\cite{SL_nordic95}]
\label{lem:precongruence}
$\preorder$ is a preorder $(${\em \ie}reflexive and transitive$)$.
\end{lemma}

Finally, we find the following characterization of $\preorder$ useful in the
algorithms we will discuss later on.

\begin{lemma}
\label{lem:preorder_characterization}
Let $L_1$ be a tree and $s_1 R s_2$
iff for every $s_1 \trans{a} \mu_1$, there exists $s_2 \trans{a} \mu_2$ with
$\mu_1 \dpreorder_R \mu_2$. Then, $R = \preorder$.
\end{lemma}

\begin{IEEEproof}[Proof Sketch]
$R \subseteq \preorder$ by Def. \ref{def:strong_simulation}. $\preorder
\subseteq R$ can be proved by induction on the {\em height} of a state of $L_1$
using Lemma \ref{lem:css}.
\end{IEEEproof}

\begin{figure}
\centering
\includegraphics[scale=1.6]{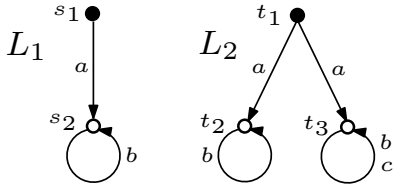}
\caption{An example showing that Lemma \ref{lem:preorder_characterization} does
not hold, in general, if $L_1$ is not a tree. Let $R = \{(s_1,t_1),
(s_2,t_2)\}$. Note that $\preorder = \{(s_1,t_1), (s_2,t_2), (s_2,t_3)\}$ and $R
\subset \preorder$.}
\label{fig:preorder_characterization_cex}
\vspace{-0.2in}
\end{figure}

Note that the condition on $R$ in the lemma is stronger than the one to make it
a strong simulation (Definition \ref{def:strong_simulation}). Also, if $L_1$ is
not a tree, we can only conclude that $R \subseteq \preorder$, in general. See
Figure \ref{fig:preorder_characterization_cex} for an example where $R \subset \preorder$.

\noindent{\bf Counterexamples to ${\bm \preorder}$.}
In the active learning problem we are interested in (Section
\ref{sec:learning_lpts}), a learner uses counterexamples to simulation
conformance as diagnostic information. We will now briefly discuss
what these counterexamples are. Let $L_1$ and $L_2$ be two LPTSes.

\begin{definition}[Language of an LPTS]
Given an LPTS $L$, we define its language, denoted $\L(L)$, as the set
$\{L' | L' ~\text{is an LPTS and } L' \preorder L\}$.
\end{definition}

\begin{lemma}
\label{lem:preorder_language}
$L_1 \preorder L_2$ iff $\L(L_1) \subseteq \L(L_2)$.
\end{lemma}

\begin{IEEEproof}
Necessity follows trivially from the transitivity of $\preorder$ and sufficiency
follows from the reflexivity of $\preorder$ which implies $L_1 \in \L(L_1)$.
\end{IEEEproof}

Thus, a counterexample $C$ can be defined as follows.

\begin{definition}[Counterexample]
\label{def:cex}
A counterexample to $L_1 \preorder L_2$ is an LPTS $C$ such that $C \in \L(L_1)
\setminus \L(L_2)$, i.e. $C \preorder L_1$ but $C \not\preorder L_2$.
\end{definition}

Now, $L_1$ itself is a trivial choice for $C$ but it does not give any more useful
information than what we had before checking the conformance.
Moreover, it is preferable to have $C$
with a special and simpler structure to efficiently work with counterexamples.
Fortunately, we have a simpler characterization using trees.


\begin{theorem}[\cite{KPC_cav12}]
\label{thm:tree_cex}
If $L_1 \not\preorder L_2$, there is a tree which serves as a counterexample.
\end{theorem}

\begin{IEEEproof}[Proof Sketch]
One can instrument the algorithm to compute the coarsest strong simulation
described earlier to obtain a tree counterexample whenever a pair of states is
removed from the current relation, making use of Lemma \ref{lem:image_based}.
\end{IEEEproof}

For example, $C_p$ in Figure \ref{fig:cex_eg}, for $p \in (0,\frac{1}{2}]$, is a counterexample to $L_1
\preorder L_2$. In another work, we showed that structures simpler than trees are not
sufficient as counterexamples, even when one of the models is
reactive~\cite{KPC_cav12}.

We note an important feature of the algorithm used to prove the above theorem~\cite{KPC_cav12}.
A counterexample $C$ generated by the algorithm is essentially a finite {\em tree execution} of
$L_1$. That is, there is a total mapping $M : S_C \to S_1$ such that for every
transition $c \trans{a} \mu_c$ of $C$, there exists $M(c) \trans{a} \mu_1$ such that $M$ restricted
to $\supp{\mu_c}$ is an injection and for every $c' \in \supp{\mu_c}$,
$\mu_c(c') = \mu_1(M(c'))$. Note that $M$ is also a strong simulation. We call
such a mapping an {\em execution mapping from $C$ to $L_1$} in the rest of the paper. An execution mapping is shown
in brackets beside the states of $C_p$ for $p = \frac{1}{2}$ in Figure \ref{fig:cex_eg}.
While our algorithm always generates
counterexamples with an {\em execution mapping}, it is possible to have a tree
counterexample, as per Definition \ref{def:cex}, without such a mapping. For
example, $C_p$ in Figure \ref{fig:cex_eg} for $p \in (0,\frac{1}{2})$ is also a
counterexample with no such {\em execution mapping}. The condition we impose on
a teacher in the active learning problem (Section \ref{sec:learning_lpts}) is
regarding this execution mapping.

%% file: learning_consistent.tex
\section{Learning a Consistent LPTS}
\label{sec:learning_consistent}

%
%
We are interested in the problem where we are given a finite
set of {\em positive} stochastic trees (\ie in the language of an
LPTS), say $\pcex$, and another finite set of {\em negative} stochastic
trees (\ie not in the language of an LPTS), say $\ncex$. These trees constitute
the samples for a learner. The goal is
to learn an LPTS $L$ such that $\pcex \subseteq \L(L)$ and $\ncex \cap
\L(L) = \emptyset$, \ie $P \preorder L$ for every $P \in \pcex$ and $N
\preorder L$ for no $N \in \ncex$. Such an $L$ is said to be {\em
  consistent} with the tree samples. Without loss of generality, assume that
$\pcex \neq \emptyset$ as otherwise, a single state LPTS with no
transitions is trivially consistent. Also, note that the LPTS obtained by merging the start
states of all trees in $\pcex$, say $L_\pcex$, trivially satisfies
$P \preorder L_\pcex$ for every $P \in \pcex$. Now, if $L$ is a consistent LPTS, it can be shown that $L_\pcex \preorder L$ and hence, by Lemma \ref{lem:precongruence},
$L_\pcex$ is also consistent. Thus, one can easily check, in polynomial time, if there
exists a consistent LPTS by checking $N \preorder L_\pcex$ for every
$N \in \ncex$. For this reason, we always assume the existence of a consistent
LPTS. Clearly, the size of $L_\pcex$ is as large as that of $\pcex$.

If possible, we would like to learn a model with the least size, or at
least have a good upper bound on its size. Such models would be
useful when automating assume-guarantee reasoning (see
Section~\ref{sec:ag}). The algorithms we propose draw
inspiration from the ones used to infer consistent non-probabilistic
automata from counterexample traces~\cite{OG_asspr92,GO_report93,CO_rairo99,GMF_fmsd08} which are based on
partitioning the state space of the counterexamples.  Let $S_\pcex =
\bigcup_{P \in \pcex} S_P$ and $S_\ncex = \bigcup_{N \in \ncex}
S_N$. First, we consider an algorithm based on the traditional state
space partitioning of $S_\pcex$. While there is an upper bound on the size of
the learnt model, we show that such partitioning is insufficient to obtain
a minimum state consistent probabilistic system (LPTS). However, as we will see
in Section \ref{sec:learning_lpts}, we find it useful in learning an unknown
target LPTS. We will then
introduce a new way of partitioning the state space, which we call
{\em stochastic} partitioning, enabling us to obtain a
minimum state consistent LPTS.


\subsection{Using State Partitions}
\label{sec:partitions}

The first algorithm uses traditional partitions of $S_\pcex$.
For a partition $\Pi$ of $S_\pcex$, let $E_\Pi$ denote the set of equivalence
classes under $\Pi$  and for a state $s \in S_\pcex$, we let
$[s]_\Pi$ denote the equivalence class of $s$ (we drop the subscript $\Pi$ when
it is clear from the context). We always 
assume that $[s^0_P]_\Pi =
[s^0_Q]_\Pi$ for every $P,Q \in \pcex$, \ie the start states of all the positive
counterexamples are mapped to the same equivalence class.

\begin{definition}[Quotient LPTS]
\label{def:quotient_lpts}
Given a partition $\Pi$ of $S_\pcex$, define the {\em quotient LPTS}, denoted
$\pcex/\Pi$, as the LPTS $\langle E_\Pi,e^0,\alpha,\tau \rangle$ where $e^0 =
[s^0_P]_\Pi$ for every $P \in \pcex$, $\alpha = \bigcup_{P \in \pcex} \alpha_P$ and
$(e,a,\mu) \in \tau$ iff there exists $(s,a,\mu_p) \in \tau_P$ for some
$P \in \pcex$ with $[s]_\Pi = e$ such that $\mu = \lift{\mu_p}$ where
$\lift{\mu_p}(e') = \sum_{s' \in e'} \mu_p(s')$ for all $e' \in E_\Pi$.
\end{definition}

It can be easily shown that a quotient is always a well-defined LPTS.
In the following, $\Pi$ is a partition of $S_\pcex$.

\begin{lemma}
\label{lem:partition_p_consistency}
$\pcex/\Pi$ is consistent with $\pcex$ for all $\Pi$.
\end{lemma}

\begin{IEEEproof}[Proof Sketch]
One can show that $\{(s,[s]_\Pi) | s \in S_P\}$ is a
strong simulation between $P$ and $\pcex/\Pi$ for every $P \in \pcex$.
\end{IEEEproof}

\begin{definition}[Consistent Partition]
\label{def:consistent_partition}
$\Pi$ is defined to be {\em consistent} iff $\pcex/\Pi$
is consistent with $\ncex$, \ie for every $N \in \ncex$, $N \not\preorder
\pcex/\Pi$.
\end{definition}

Thus, we reduce the problem of finding a consistent LPTS to that of finding a
consistent partition. As we show below, we can always find a consistent
partition with a {\em bounded size}, where the {\em size} of $\Pi$ is $|E_\Pi|$.

\begin{lemma}
\label{lem:partition_exists}
If $L$ is an LPTS of $k$ states consistent with $\pcex$, then there is a
$\Pi$ of size at most $2^k$ such that $\pcex/\Pi \preorder L$.
\end{lemma}

\begin{IEEEproof}[Proof Sketch]
Let $P \in \pcex$. As $P \preorder L$, there is a strong simulation $R_P \subseteq
S_P \times S_L$ with $s^0_P R_P s^0_L$. As $P$ is a tree,
$s^0_P$ is not in the support of any distribution and hence,
assume without loss of
generality that $R_P(s^0_P) = \{s^0_L\}$. Let $R = \bigcup_{P \in \pcex} R_P$. Now, $R$
induces a partition $\Pi$ of $S_\pcex$ such that
for $s_1, s_2 \in S_\pcex$, $[s_1]_\Pi = [s_2]_\Pi$ iff $R(s_1) = R(s_2)$. Note
that $[s^0_P]_\Pi = [s^0_Q]_\Pi$ for $P,Q \in \pcex$. The
size of $\Pi$ is clearly bounded by $2^k$. Now, we can show that
$\{([s_p]_\Pi, s_l) | s_p R s_l\}$ is a strong simulation between $\pcex/\Pi$
and $L$.
\end{IEEEproof}

Note that, if $L$ and every $P \in \pcex$ is an LTS, an upper
bound of $k$ on the size can be shown by choosing $R_P$ in the proof to be a
function.
The following is now immediate, using Lemmas \ref{lem:precongruence} and
\ref{lem:partition_p_consistency}.

\begin{corollary}
\label{cor:separating_lpts}
For every consistent LPTS of $k$ states, there is a consistent
partition of size at most $2^k$.
\end{corollary}

\begin{figure}
\centering
\includegraphics[scale=1.6]{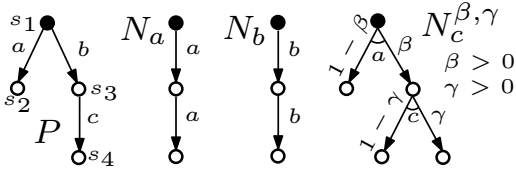}
\caption{Positive ($P$) and negative ($N_a,N_b,N^{\beta,\gamma}_c$) tree samples.}
\label{fig:partition_stochastic_cex}
\vspace{-0.2in}
\end{figure}

\begin{figure}
\centering
\includegraphics[scale=1.6]{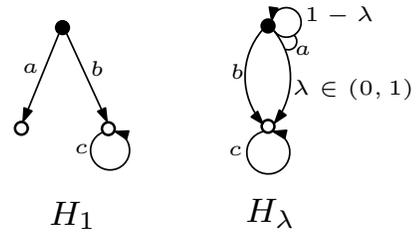}
\caption{Quotients for least size partition ($H_1$) and stochastic partition
($H_\lambda$) of $P$ in Figure \ref{fig:partition_stochastic_cex}.}
\label{fig:partition_stochastic}
\vspace{-0.2in}
\end{figure}

\noindent{\bf Observation.} This shows that if $L$ is a minimum state consistent LPTS, there exists a
consistent partition of $S_\pcex$ of size at most exponential in $|S_L|$. While
there may be a better bound, this way of partitioning $S_\pcex$ can
not guarantee a minimum state consistent LPTS in general. For example, $H_1$ in
Figure \ref{fig:partition_stochastic} is the quotient for a least sized consistent
partition of $P$ for the trees in Figure
\ref{fig:partition_stochastic_cex} (obtained by merging $s_3$ and $s_4$). On the other hand, $H_\lambda$, where $\lambda$ is any value in 
$(0,1)$, is another consistent LPTS with one less state.

\noindent{\bf Algorithm.}
A na\"{i}ve algorithm for finding a {\em least-sized consistent
partition} is to enumerate all the partitions of $S_\pcex$, with increasing
size, and for each of them, check if the corresponding quotient simulates any
tree in $\ncex$. Alternatively, we can cast it as an instance of the
satisfiability problem over linear rational arithmetic, as shown below.
In general, this is more efficient than the exhaustive search in the na\"{i}ve algorithm,
and also prepares the ground for an algorithm we discuss in the next
subsection.

First, we describe the encoding to check if there is a consistent partition of
size at most a given $k$. Let $e_i$ denote the equivalence class $i$ for $1 \le i \le k$.
For each $i$ and state $s \in S_\pcex$, we introduce a new boolean variable, say
$v_{[s]=i}$, to denote $[s]=e_i$. We add the constraint {\tt xor}$(v_{[s]=1}, \dots,
v_{[s]=k})$ for every $s \in S_\pcex$ for the partition to be well-defined.
Moreover, we fix $e_1$ to be the start state of the resulting quotient and have a
constraint that $v_{[s^0_P]=1}$ for every $P \in \pcex$ as $e_1$ should now
contain all the start states (Definition \ref{def:quotient_lpts}).

Now, to encode consistency, we want to say that no tree $N \in \ncex$ is
simulated by the resulting quotient. We can avoid introducing a universal
quantification over all possible strong simulations by finding a way to say that
$(s^0_N,e_1)$ is not in the coarsest strong simulation, for every $N \in
\ncex$. Fortunately, we can make use of Lemma
\ref{lem:preorder_characterization} to achieve exactly this. We introduce a
boolean variable $R_{s,i}$ to denote that $s \in S_\ncex$ is related to $e_i$ by the
coarsest strong simulation. Let $t_n = (s_n,a,\mu_n)$ and $t_p =
(s_p,a,\mu_p)$ be a transition of $\ncex$ and $\pcex$, respectively, on the same action
$a$, and $1 \le i \le k$. Consider the expression $d_{\mu_n,\mu_p} \land v_{[s_p]=i}$,
denoted $\sigma_{t_n,i,t_p}$. If $d_{\mu_n,\mu_p}$
denotes $\mu_n \dpreorder_R \lift{\mu_p}$, then this expression has the meaning
that $[s_p] = e_i$ and the transition corresponding to $t_p$
in the quotient, \viz $e_i \trans{a} \lift{\mu_p}$, simulates
$t_n$. If $X(s)$ denotes the set of all transitions outgoing from $s
\in S_\ncex$, $Y(a)$ denotes the set of all transitions in $\pcex$ on
action $a$ and $\act{t}$ denotes the action for the transition $t$, we
add
\[
  R_{s,i} \iff \bigand_{t_n \in X(s)} \bigor_{t_p \in
Y(\act{t_n})} \sigma_{t_n,i,t_p}
\]
according to Lemma \ref{lem:preorder_characterization}.

$\lift{\mu_p}(e_i)$ can be encoded as $\sum_{s \in \supp{\mu_p}} l_{\mu_p,i,s}$
where $l_{\mu_p,i,s}$ denotes the {\em contribution} of $s$ to the lifted
probability of $e_i$ under $\mu_p$ and satisfies
\[
  (v_{[s]=i} \implies l_{\mu_p,i,s} = \mu_p(s)) \land (\neg v_{[s]=i} \implies
l_{\mu_p,i,s} = 0).
\]
$d_{\mu_n,\mu_p}$ is encoded as follows. If we
use Definition \ref{def:weight_function_based} alone, we need to introduce a
nested existential quantifier for the weight function (to say that
$d_{\mu_n,\mu_p}$ iff {\em there is} a weight function satisfying the conditions). To avoid this nested
quantification, we also make use of Lemma \ref{lem:image_based}. First, we introduce a
variable for the weight function and encode the constraints of Definition
\ref{def:weight_function_based} if $\dpreorder_R$ holds between the
distributions. We also introduce a variable for the witness subset $S \subseteq
\supp{\mu_p}$ and encode the condition of Lemma \ref{lem:image_based} when
$\dpreorder_R$ fails to hold. This variable for the witness subset can, in turn, be
encoded using individual boolean variables for each $s \in \supp{\mu_p}$. We
also need boolean variables for the image of this witness subset under $R$. The
details are straightforward and left to the reader.
Finally, we encode consistency by having the constraint $\neg R_{s^0_N,1}$ for
every $N \in \ncex$.

It is not hard to show that the encoding is correct, \ie the resulting
encoding is satisfiable iff there is a consistent partition of size at most $k$. One can then obtain an
algorithm to find a least-sized consistent partition by starting with $k=0$ and
incrementing it as long as the encoding for $k$ is unsatisfiable. As
satisfiability over linear rational arithmetic is decidable, this is
guaranteed to terminate from Corollary \ref{cor:separating_lpts}.

\begin{theorem}
The above described algorithm to find a least-sized consistent partition of
$S_\pcex$ terminates.
\end{theorem}


\subsection{Using Stochastic Partitions}
\label{sec:stochastic_partitions}

As noted above, the quotient of a least-sized consistent partition need not have
the least number of states. We observe that the main reason for this is not
being able to partition $S_\pcex$ such that there is a one-to-one correspondence
between the equivalence classes and $S_L$, instead of the current $2^{S_L}$ for a consistent LPTS $L$ (proof of
Lemma \ref{lem:partition_exists}). This suggests that we can learn a minimum state consistent LPTS if we can find a way to group
the states of $S_\pcex$ (groups need not be disjoint) with such a
correspondence. This will then imply that if there is a minimum state consistent LPTS $L$, we can use this
grouping to obtain an equally sized consistent LPTS. One can then automate the search for
such a grouping using constraint solving.

Let $L$ be a consistent LPTS and let us see what we can do to group
$S_\pcex$ to have the above one-to-one correspondence with $S_L$. Consider
Figure \ref{fig:dist_rel_general} again and let $\mu_1$ be outgoing from the
root of some tree $P$ in $\pcex$ and $\mu_2$ appear in $L$. Let there be three
groups (initially empty), one per state
in $\supp{\mu_2}$, say $G_{t_1}$, $G_{t_2}$ and $G_{t_3}$. As explained in Section \ref{sec:prelims},
having $\mu_1 \dpreorder_R \mu_2$, for some $R$, can be thought of as finding a
way of {\em splitting} the probabilities in both the distributions and pairing
states, already in $R$, to directly match the probabilities. We would like to
use this matching to group the states of $S_\pcex$. In particular, looking at
the figure, we would like to place the two splits of $s_1$ ($s_2$) in $G_{t_1}$ and $G_{t_2}$
($G_{t_2}$ and $G_{t_3}$), respectively.

As the probability of each split of a state in $\supp{\mu_1}$ is matched with
that of some split of exactly one state in $\supp{\mu_2}$, one can also think of the above
grouping in the following alternative way. As the
probability of $\frac{1}{2}$ for $s_1$ is split into $\frac{1}{3}$ and $\frac{1}{6}$,
$s_1$ can be seen as being put in $G_{t_1}$ with probability
$\frac{1/3}{1/2} = \frac{2}{3}$ and in $G_{t_2}$ with probability
$\frac{1/6}{1/2} = \frac{1}{3}$. Thus, instead of putting $s_1$
deterministically into one group, it is put {\em stochastically} into multiple
groups. Let these splits of $s_1$ put in $G_{t_1}$ and $G_{t_2}$ be
$s_1[t_1]$ and $s_1[t_2]$, respectively.

Now, consider $s_1[t_1]$. As the corresponding probability of $\frac{1}{3}$ is
matched with that of some split of $t_1$ (implying $s_1 R t_1$), and as $s_1$ is
not in the support of any distribution other than $\mu_1$ (note that $P$ is a
tree), we need not consider if $s_1$ is related, by $R$,
to any other state in $L$, as far as $s_1[t_1]$ is concerned. And therefore, any distribution outgoing from this
split of $s_1$ will only need to be related to some distribution outgoing from $t_1$ (by
$\dpreorder_R$). Similarly, for $s_1[t_2]$ and $t_2$. Now, if $\mu_3$ is a
distribution outgoing from $s_1$ in $P$, we may want to relate it to a
distribution $\mu$ outgoing from $t_1$ (for $s_1[t_1]$) and another distribution
$\mu'$ outgoing from $t_2$ (for $s_1[t_2]$). For a state $s_3 \in \supp{\mu_3}$,
considering $\mu_3 \dpreorder_R \mu$ and $\mu_3 \dpreorder_R \mu'$ both hold,
following the above described {\em stochastic} grouping may
result in two different ways of grouping $s_3$. Thus, we need to {\em remember}
the group of its parent, denoted by $\parent{\cdot}$, when grouping a state in $S_\pcex$.

This is the main motivation behind a {\em stochastic} partition, which is
defined below.

\begin{definition}[Stochastic Partition]
\label{def:stochastic_partition}
A {\em stochastic partition} of $S_\pcex$ is a tuple $(G,\{[s]\}_{s \in S_\pcex})$ where $G \subseteq
2^{S_\pcex}$ and $[s] : G \to \dist{G}$ for every $s \in S_\pcex$, such that $\bigcup G = S_\pcex$ and 
  \begin{enumerate}
  \item there is a $g^0 \in G$ such that for every $P \in \pcex$ and $g \in G$, $[s^0_P](g) =
\dirac{g^0}$ 
and
  \item for every non-root state $s \in S_\pcex$ and $g \in G$, $[s](g)$ is defined iff
$[\parent{s}](g')(g)>0$
for some $g' \in G$.
  \end{enumerate}
Furthermore, $s \in g$ iff 
$[s](g')(g)>0$
for some
$g' \in G$, for every $s \in S_\pcex$ and $g \in G$.

We use $(G_\Pi,\{[s]_\Pi\}_s)$ for a stochastic partition $\Pi$ and when
$\Pi$ is clear, we drop the subscripts.
\end{definition}

Here, $G$ denotes the groups mentioned above and $[s]$ denotes the {\em stochastic} grouping of $s \in
S_\pcex$ given a group of its parent. Point $1$ above says that the start
states of all trees in $\pcex$ go {\em deterministically} to a designated group.
Note that the start states have no parents and the dependence of $[s^0_P]$ on
an argument is just a notational convenience.
And point $2$ says that for every non-root state $s$, $[s]$ is only defined for a {\em valid} group of its
parent. We implicitly
assume that 
$[s](g')(g)=0$ for every $g \in G$
if $[s]$ is not defined at $g'$.

Now, we define the quotient of a stochastic partition in the following way.

\begin{definition}[Quotient LPTS]
\label{def:quotient_lpts_2}
Given a stochastic partition $\Pi = (G,\{[s]\}_s)$ of $S_\pcex$, define the {\em quotient LPTS}, denoted
$\pcex/\Pi$, as the LPTS $\langle G,g^0,\alpha,\tau \rangle$ where
$g^0 \in G$ is such that $[s^0_P](g) = \dirac{g^0}$ for every
$P \in \pcex$ and $g \in G$, $\alpha = \bigcup_{P \in \pcex} \alpha_P$ and
$(g,a,\mu) \in \tau$ iff there exists $(s,a,\mu_p) \in \tau_P$, for some
$P \in \pcex$ such that $s \in g$ and
for every $g' \in G$,
\[
  \mu(g') = \sum_{s' \in g'} [s'](g)(g') \cdot \mu_p(s').
\]
We denote this relation between $\mu$ and $\mu_p$ by $\mu =
\lift{\mu_p, g}$.
\end{definition}

Thus, $(g,a,\mu) \in \tau$ iff there is a state $s \in g$ with $s \trans{a}
\mu_p$ and $\mu$ is obtained by {\em lifting}
$\mu_p$, given that $s \in g$.
For this to make sense, we need to show that the lifting is a valid
distribution.
In the following, $\Pi = (G,\{[s]\}_s)$ is a stochastic partition. 

\begin{lemma}
\label{lem:lifting_valid}
$\pcex/\Pi$ is a well-defined LPTS.
\end{lemma}

We have the following lemma analogous to classical partitions.

\begin{lemma}
\label{lem:stochastic_partition_p_consistency}
$\pcex/\Pi$ is consistent with $\pcex$ for all $\Pi$.
\end{lemma}

\begin{IEEEproof}[Proof Sketch]
One can show that $\{(s,g) | g \in G, s \in S_P \cap g\}$ is a strong simulation
between $P$ and $\pcex/\Pi$ for $P \in \pcex$.
\end{IEEEproof}


Consistency of a stochastic partition is defined in the same way as Definition
\ref{def:consistent_partition}. Thus, we reduce the problem of finding a minimum state consistent LPTS to that
of finding a {\em least-sized} consistent stochastic partition where the {\em size} of a stochastic partition
is its number of groups.

\begin{lemma}
\label{lem:stochastic_partition_exists}
If $L$ is an LPTS of $k$ states consistent with $\pcex$, then there is a
$\Pi$ of size at most $k$ with $\pcex/\Pi \preorder L$.
\end{lemma}

\begin{IEEEproof}[Proof Sketch]
Let $P \in \pcex$. As $P \preorder L$, there is a strong simulation $R_P \subseteq
S_P \times S_L$ with $s^0_P R_P s^0_L$. Let $R = \bigcup_{P \in \pcex} R_P$. Now, construct
a stochastic partition with at most $|S_L|$ many groups following the
intuitive explanation we gave when motivating stochastic partitions. For
distributions $\mu_p \in \dist{S_\pcex}$ and $\mu_l \in \dist{S_L}$, the
stochastic groupings of a state $s \in \supp{\mu_p}$ is obtained by using a
weight function showing $\mu_p \dpreorder_R \mu_l$. In particular, $s$ is put in the group
corresponding to $s_l \in S_L$ with probability $w(s,s_l)/\mu_p(s)$ where $w$ is
the weight function which is uniquely chosen given $\mu_p$ and $\mu_l$.
Moreover, $\mu_l$ and this grouping depend on the group of $\parent{s}$.
Once such a stochastic partition $\Pi$ is built, we can show
that $\{(g,s_l) | g ~\text{is the group corresponding to}~ s_l\}$ is a strong
simulation between $\pcex/\Pi$ and $L$.
\end{IEEEproof}

Our main result follows as an immediate corollary, using Lemmas
\ref{lem:precongruence} and \ref{lem:stochastic_partition_p_consistency}.

\begin{corollary}
\label{cor:separating_min_lpts}
For every consistent LPTS of $k$ states, there is a consistent stochastic
partition of size at most $k$.
\end{corollary}

So, we can obtain a minimum state consistent LPTS by constructing the quotient for
a consistent stochastic partition of $S_\pcex$ of the least size. For example,
$H_\lambda$, $\lambda \in (0,1)$, in Figure \ref{fig:partition_stochastic} is the quotient for a least sized consistent
stochastic partition for the trees in Figure \ref{fig:partition_stochastic_cex}
(where $s_1$ goes to group $1$, $s_2$ goes to group $2$ with probability
$\lambda$ and to group $1$ with $1-\lambda$ and $s_3$ and $s_4$ go to group $2$).
We describe an algorithm to find a {\em least-sized consistent stochastic
partition} by casting it as an instance of the satisfiability problem over linear
rational arithmetic.

\noindent{\bf Algorithm.} The encoding is similar to the case of partitions in
the previous subsection. To find a
stochastic partition of size at most a given $k$, let $g_i$ denote the group $i$ for $1 \le i
\le k$. Introduce a non-negative rational
variable $v_{[s](i),j}$ to denote
$[s](g_i)(g_j)$ for every $s \in S_\pcex$, $1 \le i,j \le k$. For every $i$ and
$s \in S_\pcex$, add the constraint $\left( \sum_{1
\le j \le k} v_{[s](i),j} = 1 \right) \lor \left( \sum_{1 \le j \le k}
v_{[s](i),j} = 0 \right)$ to denote
that $[s](g_i)$ is a distribution or is undefined.
Then, we encode points $1$ and $2$ of Definition
\ref{def:stochastic_partition} by adding the constraint $v_{[s^0_P](i),1}=1$ for every
$i$ and $P \in \pcex$, making $g_1$ the start state of the quotient, and adding
\[
  \sum_{1 \le j \le k} v_{[s](i),j} = 1 \iff
\sum_{1 \le l \le k} v_{[\parent{s}](l),i} > 0
\]
for every non-root state $s$ and $i$. This
ensures that the stochastic partition obtained is well-defined.

Encoding consistency is the same as before except for
$\sigma_{t_n,i,t_p}$ ($t_n$, $i$ and $t_p$ are as before)
which will now be
\[
  d_{\mu_n,\mu_p,i} \land \sum_{1 \le j \le k} v_{[s_p](j),i}>0.
\]
where $d_{\mu_n,\mu_p,i}$ denotes $\mu_n \dpreorder_R \lift{\mu_p,g_i}$.
Thus, we will check if there is a group of $\parent{s_p}$
(summation over
$1 \le j \le k$) for which $s_p \in g_i$ and $\mu_n \dpreorder_R
\lift{\mu_p,g_i}$. For a $j$, $\lift{\mu_p,g_i}(g_j)$ is encoded as $\sum_{s \in \supp{\mu_p}}
v_{[s](i),j} \cdot \mu_p(s)$. Rest of the encoding is similar.

We can similarly show the correctness of the encoding and the termination of the
algorithm follows from Corollary \ref{cor:separating_min_lpts}.

\begin{theorem}
\label{thm:min_consistent}
The problem of learning a minimum state consistent LPTS with $\pcex$ and $\ncex$ is
decidable.
\end{theorem}

%% file: learning_lpts.tex
\section{Active Learning for LPTSes}
\label{sec:learning_lpts}

We now consider the problem of learning the language of an LPTS, \ie
learning an LPTS up to simulation equivalence (following Lemma
\ref{lem:preorder_language}), in the framework of active learning.
Let $U$ be an unknown target LPTS. The learning framework has a learner and a teacher.
The goal of the learner is to learn an LPTS $L$ such that $L \simeq U$. To that
effect, the learner maintains a hypothesis LPTS $H$. The process of learning
proceeds in rounds where in each round, the learner makes a query to the teacher
and updates $H$ based on the response. For reasons mentioned in the introduction,
we only consider a single type of queries in this paper where the learner conjectures $H$ as (simulation) equivalent
to $U$. In response to such a query, the teacher is expected to check whether $H
\simeq U$ holds and otherwise, return a counterexample. If it is a
counterexample to $H \preorder U$ ($U \preorder H$), it is called a {\em
negative} ({\em positive}) counterexample. Following Section \ref{sec:prelims}, we
assume that the counterexamples are always trees. Furthermore, there should always exist an
LPTS consistent with all of the counterexamples,
\ie simulating all the positive counterexamples and none of the negative
counterexamples, received by the learner so
far. Also, every conjecture $H$ made by the learner should be consistent with
the counterexamples received so far, in the above sense.

Unfortunately, the framework, as described above, is too general to be useful,
as the following lemma shows.

\begin{theorem}
\label{thm:learning_lpts_undecidable}
The problem of learning an unknown LPTS $U$ is undecidable in the active
learning framework.
\end{theorem}

\begin{IEEEproof}[Proof Sketch]
We show that there is no algorithm to learn the unknown target $U_\lambda$,
which first performs an action $a$ and goes to a state with (unknown) probability
$\lambda$ to loop on action $b$ or goes to another state with the remaining
probability to deadlock,
by describing an adversarial teacher which manipulates the value of $\lambda$ as necessary to
keep generating counterexamples. After choosing an initial value of $\lambda$,
the teacher returns a counterexample as long as the hypothesis is not simulation
equivalent to the target. If a hypothesis simulation equivalent to the target
is conjectured, the teacher increases the value of $\lambda$ just enough to have
the new target not simulated by the hypothesis, while still being consistent with all the previously generated
counterexamples, and a new (positive) counterexample can then be generated.
\end{IEEEproof}


The main reason behind the theorem is that {\em it is not necessary} for the
positive tree counterexamples returned by the teacher to have an {\em execution
mapping} to $U$ (see Section \ref{sec:prelims}). Such a teacher can be seen as an
adversary which can choose the probability
values in the counterexamples returned, which are infinitely many, to
make the learner never converge to the desired probabilities.

But, in practice, to be able to apply the learning framework in a given setting,
one needs to implement the teacher's algorithm and we are not aware of any
algorithm to generate counterexamples other than the one discussed in Section
\ref{sec:prelims}. As mentioned before, this algorithm has an interesting property that the generated
counterexamples have an {\em execution mapping} to $L_1$ when $L_1 \preorder L_2$
fails. This suggests us to impose the following {\em friendliness} condition on a
teacher.

\begin{condition}[Friendly Teacher]
\label{cond:teacher_pos}
Every positive (negative) counterexample returned by the teacher should have an
execution mapping to $U$ ($H$).
\end{condition}

First of all, we observe that the proof of Theorem
\ref{thm:learning_lpts_undecidable}
no longer works because an update to $\lambda$ may violate
Condition $1$ on any positive counterexample already returned. In fact, as we show below,
the problem becomes decidable. Let $\pcex$ and $\ncex$ denote the sets of
positive and negative counterexamples, returned by the teacher so far,
respectively. First, consider the pseudo-code in Algorithm
\ref{algo:active_learning}. It suggests a method of using the algorithms
described in Section \ref{sec:learning_consistent} by treating $\pcex$ and
$\ncex$ as the tree samples. There is a choice at line $6$ to use partitions or
{\em stochastic} partitions.

\begin{algorithm}
\small
\caption{Active Learning Loop.}
\label{algo:active_learning}
\begin{algorithmic}[1]
\STATE $\pcex = \ncex = \emptyset$
\STATE $H \leftarrow$ single state LPTS with no transitions
\REPEAT
  \STATE conjecture $H$ to the teacher
  \STATE update $\pcex$ and $\ncex$ from returned counterexamples, or exit
  \STATE obtain a least sized consistent (stochastic) partition $\Pi$
  \STATE $H \leftarrow \pcex/\Pi$
\UNTIL{{\tt false}}
\end{algorithmic}
\end{algorithm}

First, we show that using traditional partitions at line $6$ makes the problem of learning a target
decidable.

%
%

\begin{lemma}
\label{lem:learning_termination}
The active learning loop of Algorithm \ref{algo:active_learning} terminates
under Condition \ref{cond:teacher_pos} on the teacher and using partitions at
line $6$ with the number of states of each intermediate hypothesis $H$ bounded
by that of $U$.
\end{lemma}

\begin{IEEEproof}[Proof Sketch]
Consider an arbitrary iteration of the learning loop.
First of all, due to Condition \ref{cond:teacher_pos}, the quotient of the partition induced
by the execution mappings from the positive counterexamples to $U$ is a
{\em sub-structure} of $U$ and hence, is trivially simulated by $U$ and is a consistent LPTS. As the
algorithm finds a {\em least-sized} consistent partition, its size is
bounded by $|S_U|$.

Then, notice that every future hypothesis is consistent with any new
counterexample returned, and hence, is distinct from the current one.
Moreover, due again to Condition \ref{cond:teacher_pos}, and as {\em lift} only
adds probabilities, one can show that there are only finitely many possible
distributions for a given partition size.

We conclude that the algorithm terminates.
\end{IEEEproof}

Thus, we have the following result.

\begin{theorem}
\label{thm:learning_lpts_cond1_decidable}
The problem of learning an unknown LPTS is decidable in the active learning
framework, with Condition \ref{cond:teacher_pos} on the teacher.
\end{theorem}

It is sometimes desirable to learn an LPTS with the least number of states.
While the algorithm described above learns an LPTS, it is not
guaranteed to output a minimum state LPTS simply because each hypothesis need
not have the least number of states (see Section \ref{sec:partitions}).
This suggests us to impose the following condition on the learner.

%
%

\begin{condition}[Learner]
\label{cond:learner}
Every hypothesis $H$ made by the learner is a minimum state LPTS consistent with
$\pcex$ and $\ncex$.
\end{condition}

If there is a learning algorithm under Conditions \ref{cond:teacher_pos} and
\ref{cond:learner}, then it is guaranteed to
output a minimum state LPTS which is (simulation) equivalent to $U$. But,
there is no such algorithm as we show below.

\begin{theorem}
\label{thm:learning_min_lpts_undecidable}
The problem of learning an unknown LPTS $U$ is undecidable in the active
learning framework, with both Condition \ref{cond:teacher_pos} on the teacher and
Condition \ref{cond:learner} on the learner.
\end{theorem}

\begin{IEEEproof}[Proof Sketch]
We show that there is no algorithm to learn (unknown) $H_1$ in Figure
\ref{fig:partition_stochastic}, by describing an
adversarial teacher which can return a counterexample for any conjectured
hypothesis. Initially, the teacher keeps returning negative counterexamples, if
there are transitions on actions other than $a$, $b$ and $c$ in the hypothesis, or
the positive counterexample $P$ in Figure~\ref{fig:partition_stochastic_cex}
until the learner conjectures a single-state LPTS with
self-loops on these three actions.
Thereafter, if a conjectured hypothesis has transitions on only $a$, $b$ and $c$ and simulates
$P$, the teacher returns $N_a$ 
to force the future hypotheses to have
at least two states and in every future round, returns $N_b$ or
$N^{\beta,\gamma}_c$ in the figure, as necessary. One can show that there are
always suitable values of $\beta$ and $\gamma$ whenever $N^{\beta,\gamma}_c$ needs to
be returned and the learner always conjectures a two state LPTS. In fact,
$H_\lambda$ is always a consistent LPTS for a suitable $\lambda \in (0,1)$.
\end{IEEEproof}

%

However, we obtain a semi-algorithm to the problem by using {\em stochastic}
partitions at line $6$ of Algorithm \ref{algo:active_learning}. That is, if the
algorithm terminates, it is guaranteed to learn the target with the least number of
states. Correctness is immediate from Theorem \ref{thm:min_consistent}.


%% file: learning_assumptions.tex
\section{Learning Assumptions for\\Compositional Reasoning}
\label{sec:ag}

As mentioned in the introduction, the original motivation for this
work was to automate assume-guarantee style reasoning for simulation
conformance. Assume-guarantee reasoning~\cite{Pnueli_lmcs85} is a
compositional technique that breaks up the verification of large
systems into that of its components for increased scalability. When
checking individual components, the method uses assumptions about
their environments and {\em discharges} them on the rest of the system. For
a system of two components, such reasoning is captured by the
following simple assume-guarantee rule ({\sc ASym}).

  \begin{mathpar}
  \inferrule[]
            {L_1 \parallel A \preorder P \\ L_2 \preorder A}
            {L_1 \parallel L_2 \preorder P}
  \end{mathpar}

Several other assume-guarantee rules have been proposed, some of them
involving symmetric~\cite{PGB+_fmsd08} or circular reasoning~\cite{AHJ_concur01,PGB+_fmsd08,KNP+_tacas10}.
Despite its simplicity,
rule {\sc ASym} has been proven most effective in practice and has
been studied extensively mainly in a non-probabilistic setting, for
different notions of conformance~\cite{PGB+_fmsd08,CCS+_cav05,FKP_fase11}.

In our case, $L_1$, $L_2$, $A$ and $P$ are LPTSes with $P$ standing
for the {\em specification} which the composition $L_1 \parallel L_2$
should conform to, where $\parallel$ is defined below.

\begin{definition}[Composition~\cite{SL_nordic95}]
\label{def:composition}
The parallel composition of $L_1$ and $L_2$, denoted $L_1 \parallel L_2$, is
defined as the LPTS $\langle S_1 \times S_2, (s^0_1,s^0_2),
\alpha_1 \cup \alpha_2, \tau \rangle$ where $(s_1,s_2) \trans{a} \mu$ iff
  \begin{enumerate}
  \item $s_1 \trans{a} \mu_1$, $s_2 \trans{a} \mu_2$ and $\mu =
\mu_1 \product \mu_2$, or
  \item $s_1 \trans{a} \mu_1$, $a \not\in \alpha_2$ and $\mu = \mu_1
\product \dirac{s_2}$, or
  \item $a \not\in \alpha_1$, $s_2 \trans{a} \mu_2$ and $\mu = \dirac{s_1}
\product \mu_2$. 
  \end{enumerate}
Here $\nu_1 \product \nu_2 \in \dist{S_1 \times S_2}$, such that $\nu_1 \product
\nu_2 : (s_1,s_2) \mapsto \nu_1(s_1) \cdot \nu_2(s_2)$, for $\nu_1 \in \dist{S_1}, \nu_2
\in \dist{S_2}$.
\end{definition}

The main challenge in using assume-guarantee reasoning is to
automatically come up with a {\em small} assumption $A$ satisfying the premises. We
first note that the proposed rule is sound and complete~\cite{KPC_cav12}. Completeness, obtained trivially by replacing $A$
with $L_2$, is essential to guarantee termination of
our proposed algorithm.  Previous attempts at automating
assume-guarantee reasoning using learning in a probabilistic setting
have been restricted to checking probabilistic reachability properties
using either an incomplete rule~\cite{FKP_fase11} or algorithms which
may not terminate~\cite{FHK+_atva11}.

Motivated by the success of existing applications of active learning
to assume-guarantee reasoning~\cite{PGB+_fmsd08,CCS+_cav05,CFC+_tacas09}, we propose to use the
active learning framework presented in Section \ref{sec:learning_lpts}
to learn an intermediate assumption $A$ in the rule {\sc ASym}.
We describe an algorithm for the problem using learning and show termination below.

\noindent{\bf Teacher.}
The teacher is implemented by two conformance checks corresponding to
the two premises of the rule, checked in any order.

\begin{itemize}
\item {\em Premise 1} guides the learner towards a conjecture that makes  
$L_1 \parallel A \preorder P$ true.
\item {\em Premise 2} guides the learner towards a conjecture that 
is {\em discharged} on $L_2$,
\ie that makes $L_2 \preorder A$ true.
\end{itemize}

If the conjectured $A$ satisfies both the premises, soundness of {\sc ASym}
implies $L_1 \parallel L_2 \preorder P$ holds, and the teacher
returns {\em true}.  If one of the premises fails, the teacher generates
counterexamples with an {\em execution mapping} (Section
\ref{sec:prelims}). Thus, the teacher satisfies Condition
\ref{cond:teacher_pos}. When premise $2$ fails, a {\em positive}
counterexample is returned to the learner. When premise $1$ fails, the
obtained counterexample is first {\em projected} onto $A$ and then
returned as a {\em negative} counterexample. As a counterexample $C$
to premise $1$ has an execution mapping to $L_1 \parallel A$, the
projection onto $A$ is simply the {\em contribution} of $A$ towards
$C$ in the composition. To enable this, additional information
regarding individual distributions is maintained during composition~\cite{KPC_cav12}.


\noindent{\bf Spuriousness Check.}
Note that if $L_1 \parallel L_2 \not\preorder P$, no assumption satisfies both the
premises of {\sc ASym} (violating the assumption on the existence of a
consistent LPTS in Section \ref{sec:learning_consistent}). To detect this, the learner needs to check if a
counterexample returned by the teacher exposes the failure of the conclusion of
{\sc ASym}. A {\em real} counterexample would imply that the
specification will not hold of the original system while a {\em spurious}
one would need the learner to revise its hypothesis for the
assumption. We restrict spuriousness check to {\em negative
counterexamples} following previous approaches~\cite{PGB+_fmsd08}. A
simple way is to check $N \preorder L_2$ for a negative
counterexample $N$. $N$ is real if the check succeeds and spurious,
otherwise. A slightly more involved, but practical, way is described
elsewhere~\cite{KPC_cav12}.


\noindent{\bf Algorithm.}
Now, the learner can simply use Algorithm \ref{algo:active_learning},
using {\em partitions}, to learn an intermediate assumption. As the positive
(negative) counterexamples have execution mapping to $L_2$ (A), it is
as if the unknown target is $L_2$. Note that if $P$ holds of the
system, $L_2$ is clearly an assumption satisfying the
premises. However, the algorithm is expected to terminate with a smaller assumption in
practice, which also satisfies the premises. If $P$ does not hold, the algorithm terminates with a real
counterexample. Termination is guaranteed by Lemma
\ref{lem:learning_termination}.
If we also impose Condition \ref{cond:learner}, the learner uses {\em
stochastic} partitions in Algorithm \ref{algo:active_learning} giving a semi-algorithm.

\noindent{\bf Complexity Analysis.} 
Let us now analyze the complexity of assume-guarantee reasoning using
the learning algorithm described above (with partitions). The
complexity of checking $L_1 \parallel L_2 \preorder P$ directly is
$O(\poly{|L_1| \cdot |L_2|,|P|})$, where $|L|$ denotes {\tt
max}$(|S_L|,|\tau_L|)$.

Let $d = |\tau_2|$ and $b$ be the maximum size
of the support of a distribution in $L_2$. Given a state of a candidate
assumption of size $k$ and a distribution of $L_2$, there can be at most $k^b$-many
corresponding distributions (due to non-determinism) from that state. For $k$ states and $d$
distributions, this gives a total
of $d k^{b+1}$. Therefore, there are $2^{d k^{b+1}}$ different possible candidates
of size $k$ to consider.
The total number of iterations of the learning algorithm is then bounded by
$\sum_{k=1}^{m} 2^{d k^{b+1}} = O(m 2^{d m^{b+1}})$, where $m$ is the number of
states in the final assumption output by the algorithm.

At each iteration, in the worst-case, the algorithm enumerates all the
candidate assumptions of the current size $k$ and performs simulation checks
with all the negative counterexamples. These checks have a complexity of
$O(\poly{|A|,|\ncex|,l})$, where $A$ is the final assumption, $\ncex$ is the final set of
negative counterexamples and $l$ is the largest
$|N|$, for any $N \in \ncex$. Thus, the total worst-case complexity of
the learning algorithm for computing the final assumption is
$O(\poly{|A|,|\ncex|,l} \cdot m 2^{d m^{b+1}})$.
Furthermore, the complexity of checking the two premises of {\sc ASym} is
$O(\poly{|L_1| \cdot |A|, |P|} + \poly{|L_2|,|P|})$ at every iteration.
We observe that in practice, if the assumption is small (\ie $|A| \ll
|L_2|$) this approach can be better than checking $L_1 \parallel L_2$
directly.
In other cases, however, we
would need better algorithms 
to address the problem. We leave this for future work.

%% file: conclusion.tex
\section{Conclusion}
We have presented algorithms and decidability results for the problem
of learning non-deterministic LPTSes from stochastic tree samples, using
traditional and stochastic state-space partitioning. We have also
described the application of the algorithms to automating the
discovery of assumptions for the compositional verification of LPTSes.

In the future, we would like to investigate further conditions on the
teacher that will make the active learning problem with
stochastic partitions decidable. 
We also plan to investigate the use of weak simulation for the conformance relation, 
as this will result in smaller assumptions for compositional verification.
However, algorithms for checking weak simulation are not currently known. 
Finally we plan to investigate new applications for our algorithms in learning abstractions or active model checking
and in domains other than verification.



%% file: ack_arxiv.tex
\section*{Acknowledgments}
We thank
Christel Baier,
Rohit Chadha,
Sagar Chaki,
Lu Feng,
Holger Hermanns,
Marta Kwiatkowska,
Joel Ouaknine,
David Parker,
Nishant Sinha,
Frits Vaandrager,
Mahesh Viswanathan,
James Worrell and
Lijun Zhang
for answering our questions related to
this research and the reviewers for their suggestions.

%% file: main_arxiv.bbl

%% file: appendix.tex
\subsection{Proof of Lemma \ref{lem:css}}
By Definition \ref{def:strong_simulation}, $\preorder$ is the union of all
strong simulations. It can easily be shown that union of two strong simulations is a strong
simulation and hence $\preorder$ is a strong simulation. It is also the coarsest
as it includes any strong simulation.
\IEEEQED

\subsection{Proof of Lemma \ref{lem:preorder_characterization}}
It suffices to show that $R \subseteq \preorder$ and $\preorder \subseteq R$.

$R$ is clearly a strong simulation which, by Lemma \ref{lem:css}, implies $R \subseteq \preorder$.

To prove the other direction, let $s_1 \preorder s_2$. We show that $s_1 R s_2$
by induction on the {\em height} of $s_1$ in the tree $L_1$, where the height of
a leaf state is defined to be $0$ and the height of any other state is defined
to be one plus the maximum height of any state in the support of any outgoing
distribution from that state.

For the base case, let $s_1$ be any leaf state. As $s_1$ has no outgoing
transitions, $s_1 R s_2$ trivially holds by the assumption on $R$.

For the inductive case, let the height of $s_1$ be non-zero and let $s_1 \trans{a}
\mu_1$. Then, as $\preorder$ is a
strong simulation (Lemma \ref{lem:css}), there exists
$\mu_2$ with $s_2 \trans{a} \mu_2$ such that $\mu_1
\dpreorder_\preorder \mu_2$. Let $S \subseteq \supp{\mu_1}$. We then have $\mu_1(S)
\leq \mu_2(\preorder(S))$. As every state in $\supp{\mu_1}$, and hence in $S$, has
a smaller height than that of $s_1$, by induction hypothesis, $\preorder(S)
\subseteq R(S)$ and therefore, $\mu_1(S) \leq \mu_2(R(S))$. As $S$ is
arbitrary, we conclude that $\mu_1 \dpreorder_R \mu_2$. By the assumption on
$R$, we conclude that $s_1 R s_2$.

Thus, by induction, we conclude that $\preorder \subseteq R$.
\IEEEQED

\subsection{Proof of Lemma \ref{lem:partition_exists}}
Let $P \in \pcex$. As $P \preorder L$, there is a strong simulation $R_P \subseteq
S_P \times S_L$ with $s^0_P R_P s^0_L$. As $P$ is a tree, $s^0_P$ is not in the
support of any distribution and hence, assume without loss of
generality that $R_P(s^0_P) = \{s^0_L\}$. Let $R = \bigcup_{P \in \pcex} R_P$. Now, $R$
induces a partition $\Pi$ of $S_\pcex$ such that
for $s_1, s_2 \in S_\pcex$, $[s_1]_\Pi = [s_2]_\Pi$ iff $R(s_1) = R(s_2)$. Note
that $[s^0_P]_\Pi = [s^0_Q]_\Pi$ for $P,Q \in \pcex$, satisfying the assumption
on $\Pi$ in Definition \ref{def:quotient_lpts}. The
size of $\Pi$ is clearly bounded by $2^k$.

We first show that the relation
$R' = \{([s_p]_\Pi, s_l) | s_p R s_l\}$ is a strong simulation. Let $e R'
s_l$ and $e \trans{a} \mu$. By Definition \ref{def:quotient_lpts}, there
exists $s_p \in S_\pcex$ and $\mu_p \in \dist{S_\pcex}$ with $[s_p]_\Pi = e$,
$s_p \trans{a} \mu_p$ and $\mu(e') = \sum_{s' \in e'} \mu_p(s')$ for all $e' \in E$. By the definition of
$R'$ and $\Pi$, $R(s_1) = R(s_2)$ for all $s_1,s_2 \in e$ and hence, $s_p R
s_l$. As $R$ is the disjoint union of strong
simulations, there exists $\mu_l \in \dist{S_L}$
such that $s_l \trans{a} \mu_l$ and $\mu_p \dpreorder_R \mu_l$. Let $E'
\subseteq \supp{\mu}$. Now, $\mu(E')$

  \begin{align*}
  = & \sum_{e' \in E'} \mu(e') \\
  = & \sum_{e' \in E'} \mu_p(\{s \in S_\pcex | [s]_\Pi = e'\}) & \{\text{choice
of $\mu$}\} \\
  = & \mu_p(\{s \in S_\pcex | [s]_\Pi \in E'\}) \\
  \le & \mu_l(R(\{s \in S_\pcex | [s]_\Pi \in E'\})) & \{\mu_p \dpreorder_R \mu_l\} \\
  = & \mu_l(\bigcup_{e' \in E'} R(\{s \in S_\pcex | [s]_\Pi = e'\})) \\
  = & \mu_l(\bigcup_{e' \in E'} R'(e')) & \{\text{Def. of $R'$}\} \\
  = & \mu_l(R'(E')).
  \end{align*}

So, by Lemma \ref{lem:image_based}, $\mu \dpreorder_{R'} \mu_l$. We conclude that $R'$ is a strong
simulation. For an arbitrary $P \in \pcex$, as $s^0_P R s^0_L$ and as $s^0_{\pcex/\Pi} =
[s^0_P]_\Pi$ (Definition \ref{def:quotient_lpts}), $s^0_{\pcex/\Pi} R' s^0_L$.
Therefore, $\pcex/\Pi \preorder L$.
%
\IEEEQED

\subsection{Proof of Lemma \ref{lem:lifting_valid}}
Let $(g,a,\mu) \in \tau_{\pcex/\Pi}$ be arbitrary. It suffices to show that
$\mu \in \dist{G}$. This immediately implies that $\pcex/\Pi$ is an LPTS, according to
Definition \ref{def:lpts}. Let $(s,a,\mu_p) \in \tau_P$ for some $P
\in \pcex$ such that $s \in g$ and $\mu = \lift{\mu_p,g}$ as in Definition \ref{def:quotient_lpts_2}. Now, $\sum_{g' \in G} \mu(g') =$
  \begin{align*}
  & = \sum_{g' \in G} \sum_{s' \in g'}
[s'](g)(g') \cdot \mu_p(s') \\
  & = \sum_{s' \in S_\pcex} \left( \mu_p(s') \cdot \sum_{g' : s' \in g'}
[s'](g)(g') \right) \\
  & = \sum_{s' \in S_\pcex} \left( \mu_p(s') \cdot \sum_{g' : [s'](g)(g')>0}
[s'](g)(g') \right) \quad \{\text{Definition \ref{def:stochastic_partition}}\} \\
  & = \sum_{s' \in S_\pcex} \mu_p(s') \quad \{[s'](g) \in \dist{G}\}\\
  & = 1 \quad \{\mu_p \in \dist{S_\pcex}\}.
  \end{align*}
\IEEEQED

\subsection{Proof of Lemma \ref{lem:stochastic_partition_p_consistency}}
Let $P \in \pcex$. We first show that the relation $R = \{(s,g) | g \in
G, s \in S_P \cap g\}$ is a strong simulation.

Let $s R g$ and $s \trans{a} \mu_p$. As $s \in g$, by Definition
\ref{def:quotient_lpts_2}, $g \trans{a} \mu$ where for every $g' \in G$,
\[
\mu(g') = \sum_{s' \in g'}
[s'](g)(g') \cdot \mu_p(s').
\]

It suffices to show that $\mu_p \dpreorder_R \mu$. Let $S \subseteq
\supp{\mu_p}$.
Now, $\mu_p(S)$

  \begin{align*}
  & = \sum_{s' \in S} \mu_p(s') \\
  & = \sum_{s' \in S} \sum_{g' : s' \in g'} [s'](g)(g') \cdot \mu_p(s')
\quad \{[s'](g) \in \dist{G}\} \\
  & = \sum_{g' \in G} \sum_{s' \in S \cap g'} [s'](g)(g')
\cdot \mu_p(s') \\
  & = \sum_{g' \in R(S)} \sum_{s' \in S \cap g'} [s'](g)(g')
\cdot \mu_p(s') \\
  & \quad \{\text{definition of $R$}\} \\
  & \le \sum_{g' \in R(S)} \sum_{s' \in g'}
[s'](g)(g') \cdot \mu_p(s') \\
  & = \sum_{g' \in R(S)} \mu(g') \quad \{\text{choice of $\mu$}\} \\
  & = \mu(R(S)) \\
  \end{align*}

So, by Lemma \ref{lem:image_based}, $\mu_p \dpreorder_R \mu$.
We conclude that $R$ is a strong simulation. From Definitions
\ref{def:stochastic_partition} and \ref{def:quotient_lpts_2}, $s^0_P \in
s^0_{\pcex/\Pi}$ and hence, $s^0_P R s^0_{\pcex/\Pi}$. Therefore, $P \preorder
\pcex/\Pi$.
\IEEEQED

\subsection{Proof of Lemma \ref{lem:stochastic_partition_exists}}
Let $P \in \pcex$. As $P \preorder L$, there is a strong simulation $R_P
\subseteq S_P \times S_L$ with $s^0_P R_P s^0_L$. Let $R = \bigcup_{P \in \pcex}
R_P$. For $s R s_l$ and $s \trans{a} \mu_p$, there can be one or more
transitions $s_l \trans{a} \mu_l$ with $\mu_p \dpreorder_R \mu_l$. We assume that we can always choose a
unique $s_l \trans{a} \mu_l$ with $\mu_p \dpreorder_R \mu_l$ (say, by ordering the
possible transitions in some way and choosing the first) and also that we can
always choose a unique weight function $w$ satisfying the conditions of
Definition \ref{def:weight_function_based} for $\mu_p \dpreorder_R \mu_l$.

Create a group of states of $S_\pcex$ for each $s_l \in S_L$, say $\gamma(s_l)$,
initialized to $\emptyset$ and let $\Gamma$ be the set of all these groups.
We will populate these groups by induction on the depth of a state in
$S_\pcex$ with $s \in \gamma(s_l)$ implying $s R s_l$. We will also define $\varphi(s) : \Gamma \to
\dist{\Gamma}$ for each $s \in S_\pcex$ by the same induction.
Let $s \in S_\pcex$ be arbitrary. We proceed by induction on $d(s)$, the depth of $s$.

The base case is when $d(s) = 0$ implying $s$ is a start state. $s$ is added to
$\gamma(s^0_L)$ and $\varphi(s)$ maps every $g \in \Gamma$ to $\dirac{\gamma(s^0_L)}$. Clearly, $s R
s^0_L$ and $\varphi(s)(g)(\gamma(s^0_L))>0$ for every $g \in \Gamma$.

For the inductive step, $d(s) > 0$ and let $g \in \Gamma$.
If $\parent{s} \not\in g$, $\varphi(s)(g)$ is undefined. Otherwise, let $s_l \in
S_L$ be the unique state satisfying $g = \gamma(s_l)$. Thus, $\parent{s} \in
\gamma(s_l)$ and by induction hypothesis, $\parent{s} R s_l$.
Let $\parent{s} \trans{a} \mu_p$ be the unique
transition with $s \in \supp{\mu_p}$ (as $\parent{s}$ is unique). As $R$ is the
disjoint union of strong simulations, choose $s_l \trans{a} \mu_l$ with
$\mu_p \dpreorder_R \mu_l$ as mentioned in the beginning in a unique way.
Furthermore, let $w$ be the uniquely chosen weight function satisfying the
conditions in Definition \ref{def:weight_function_based} for $\mu_p \dpreorder_R
\mu_l$. For every $s'_l \in S_L$ with $w(s,s'_l) > 0$, define
$\varphi(s)(g)(\gamma(s'_l)) = w(s,s'_l)/\mu_p(s)$ and add $s$
to $\gamma(s'_l)$. Now, $w(s,s'_l) > 0$ implies $s R s'_l$ by Definition
\ref{def:weight_function_based}. The definition also says that $\sum_{s'_l \in
S_L} w(s,s'_l) = \mu_p(s)$ which implies that $\varphi(s)(g) \in \dist{\Gamma}$.
Clearly, $\varphi(s)(g)(\gamma(s'_l))>0$.

That completes populating $\Gamma$ and defining $\varphi(s)$ for every state $s
\in S_\pcex$. Note that if $\varphi(s)(g)$ is defined, then $\parent{s} \in g$ from
the above construction and hence, $g$ is non-empty. Furthermore, every group $g$ in
$\supp{\varphi(s)(g)}$ contains $s$, again from the construction above, and hence, is non-empty.

Now, define a stochastic partition $\Pi = (G,\{[s]\}_{s \in S_\pcex})$ with $G$ containing all
the non-empty groups of $\Gamma$ and $[s]$ given by $\varphi(s)$. It is not
difficult to see that $\Pi$ is well-defined according to Definition
\ref{def:stochastic_partition}. First of all, one can easily show, using the
same induction above, that every state is added to some group and hence
$\bigcup G = S_\pcex$. Then, as discussed
above, $\varphi(s)$ is only defined for groups in $G$ and the support of any distribution in the
range set of $\varphi(s)$ is contained in $G$ and hence, $\varphi(s) : G \to \dist{G}$.
$\gamma(s^0_L)$ is the $g^0$ in Definition \ref{def:stochastic_partition}. Also,
from the way we populated groups in $G$, the condition that $s \in g$ iff there
exists $g' \in G$ such that $[s](g')(g)>0$ 
follows for every $s \in S_\pcex$ and $g \in G$.

We will now show that $\pcex/\Pi \preorder L$ by first proving that $R' = \{(g,s_l) | g \in G, g =
\gamma(s_l)\}$ is a strong simulation. Let $g R' s_l$ and $g \trans{a} \mu$. By
Definition \ref{def:quotient_lpts_2}, there exists $s \trans{a} \mu_p$ in some
$P \in \pcex$ with $s \in g$ such that for every $g' \in G$,
\[
\mu(g') = \sum_{s' \in g'} [s'](g)(g') \cdot \mu_p(s').
\]
By definition of $R'$, $g = \gamma(s_l)$ and hence, $s \in \gamma(s_l)$. From the above construction
of $\Pi$, we can then infer $s R s_l$. Now, choose $s_l \trans{a} \mu_l$ with
$\mu_p \dpreorder_R \mu_l$ as mentioned in the beginning in a unique way.
It suffices to show that $\mu \dpreorder_{R'} \mu_l$. Let $w$ be the uniquely
chosen weight function to show that $\mu_p \dpreorder_R \mu_l$.

Let $\gamma(s'_l) \in \supp{\mu}$. Then, $\mu(\gamma(s'_l))$
  \begin{align*}
  & = \sum_{s' \in \gamma(s'_l)}
[s'](g)(\gamma(s'_l)) \cdot \mu_p(s') \\
  & \quad \quad \{\text{choice of $\mu$ above}\}\\
  & = \sum_{s' \in \supp{\mu_p} \cap \gamma(s'_l)}
[s'](g)(\gamma(s'_l)) \cdot \mu_p(s') \\
  & = \sum_{s' \in \supp{\mu_p}} w(s',s'_l) \\
  & \quad \quad \{\text{from the above construction of $[s']$}\} \\
  & = \mu_l(s'_l) \\
  & \quad \quad \{\text{Definition \ref{def:weight_function_based}}\} \\
  \end{align*}

So, $\mu(g') = \mu_l(R'(g'))$ for every $g' \in \supp{\mu}$. As $R'$ maps distinct groups in $G$ to distinct
states of $S_L$, it follows that $\mu \dpreorder_{R'} \mu_l$ (by exhibiting the
trivial weight function). We conclude that $R'$ is a strong simulation. Clearly,
$s^0_{\pcex/\Pi} = \gamma(s^0_L) R' s^0_L$. Therefore, $\pcex/\Pi \preorder L$. Also, $|G| \le |S_L| = k$.
\IEEEQED

\subsection{Proof of Theorem \ref{thm:learning_lpts_undecidable}}
\begin{figure}
\centering
\includegraphics[scale=1.5]{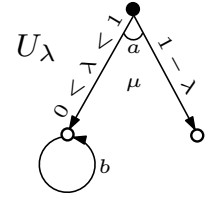}
\caption{There is no learner for the target $U_\lambda$ in presence of an unrestricted teacher.}
\label{fig:learning_lpts_undecidable}
\end{figure}

We give an example where it is impossible for the learner to converge to the
unknown target, up to $\simeq$, in presence of an adversarial teacher.

Consider $U_\lambda$ in Figure \ref{fig:learning_lpts_undecidable} where $\lambda \in (0,1)$.
For a fixed $\lambda$, $U_\lambda$ is an LPTS with the alphabet $\{a,b\}$. The
strategy for an adversarial teacher is described in Algorithm
\ref{algo:learning_lpts_undecidable} which is briefly summarized in words below. Let
$U_\lambda$, for some unknown $\lambda$, be the unknown target and $H_n$ be the hypothesis at the beginning
of every round $n \ge 1$ of the active learning loop (we count rounds beginning
with $1$). The teacher acts as an adversary by manipulating the value of
$\lambda$ as necessary and it suffices to show that there is {\em some} LPTS
consistent with all the counterexamples generated so far. So, let $\lambda_n$ be
the value of $\lambda$ at the beginning of round $n$ and let $\mu_n$ be the
corresponding distribution on $a$.

In every round $n$, the teacher first checks $H_n \preorder U_\lambda$, returning
a {\em negative} counterexample if it fails, and then checks $U_\lambda \preorder
H_n$, returning a {\em positive} counterexample if it fails. If both checks
succeed (\ie $U_\lambda \simeq H_n$), the teacher modifies the value of $\lambda$ such that $U_{\lambda_n}
\preorder U_\lambda$ but not the other way around. This is achieved by
incrementing its value at line $15$, where $\text{{\em Dist}}_a[\ncex]$ is the set
of distributions labeled by $a$ in $\ncex$. First, it computes $\lambda^+$ which is
the least of all $p^\mu_b$'s, greater than $\lambda$, and $1$ where $\mu$ is any
distribution appearing in a transition of any negative counterexample labeled by
$a$ and $p^\mu_b$ is the measure, under $\mu$, of all the states having a transition on
$b$. It then updates $\lambda$ to the mean of $\lambda$ and $\lambda^+$, \ie
$\lambda_{n+1} = (\lambda_n + \lambda^+_n)/2$.
After this update, as $\lambda > \lambda_n$, $U_{\lambda_n} \preorder U_\lambda$
holds but $U_\lambda \not\preorder U_{\lambda_n}$ and hence, $U_\lambda
\not\preorder H_n$. This ensures that a {\em positive} counterexample $P$ always
exists, justifying line $16$.

Now, it is easy to see that $\lambda^+$ at line $14$ is well-defined and
always exists. Thus, the teacher can return a counterexample for every
hypothesis made by the learner. 

We will now show that $U_{\lambda_n}$ is
consistent with $\pcex$ and $\ncex$ at the beginning of each round $n \ge 1$ by
induction on $n$, where $\pcex$ and $\ncex$ are the sets of positive and
negative counterexamples, respectively. For $n=1$, $\pcex \cup \ncex = \emptyset$ and hence,
$U_{\lambda_1}$ is consistent.

Assume that $U_{\lambda_m}$ is consistent with $\pcex \cup \ncex$ for some $m \ge 1$.
If a negative (positive) counterexample $N$ ($P$) is added to $\ncex$
($\pcex$) at line $7$ ($11$), $N \not\preorder U_{\lambda_m}$ ($P \preorder
U_{\lambda_m}$) by Definition \ref{def:cex}. As $U_{\lambda_m} =
U_{\lambda_{m+1}}$, $U_{\lambda_{m+1}}$ is consistent with $\pcex$ and $\ncex$.
Now, let $P$ be a positive counterexample added to
$\pcex$ at line $17$. Clearly, $P \preorder U_{\lambda_{m+1}}$ by Definition
\ref{def:cex}. Also, by induction
hypothesis, for every $P' \in \pcex \setminus \{P\}$, $P' \preorder
U_{\lambda_m}$ and as $U_{\lambda_m} \preorder U_{\lambda_{m+1}}$ (from above),
we obtain $P' \preorder U_{\lambda_{m+1}}$ from Lemma \ref{lem:precongruence}. Let $N \in
\ncex$. By induction hypothesis, $N \not\preorder U_{\lambda_m}$ and we need to
show that $N \not\preorder U_{\lambda_{m+1}}$. For the sake
of contradiction, assume that $N \preorder U_{\lambda_{m+1}}$.

Now, every transition outgoing from $s^0_N$ is labeled by $a$, as the only
transition outgoing from the start state of $U_{\lambda_{m+1}}$ is labeled by
$a$. Let $s^0_N \trans{a} \nu$. So, $\nu \dpreorder_\preorder \mu_{m+1}$. No state in $\supp{\nu}$ has a transition
labeled by an action other than $b$, as otherwise, $\nu \not\dpreorder_\preorder
\mu_{m+1}$. That is, every state $s$ in $\supp{\nu}$ either has no outgoing transition or has a
transition labeled by $b$. One can easily argue that this is also the case for
any transition outgoing from $s$ and so on. Consider $p^\nu_b$, the measure of all the
states having a transition on $b$ under $\nu$. We have that $p^\nu_b \le
\lambda_{m+1}$, as otherwise, $\nu \not\dpreorder_\preorder \mu_{m+1}$.

If $p^\nu_b \le \lambda_m$, clearly $N \preorder U_{\lambda_m}$ which leads to a
contradiction. So, $p^\nu_b > \lambda_m$. But then, by construction of
$U_{\lambda_{m+1}}$ (line $14$ of Algorithm
\ref{algo:learning_lpts_undecidable}), $\lambda_{m+1} < p^\nu_b$ leading to a contradiction.

We conclude that $N \not\preorder U_{\lambda_{m+1}}$. This completes the
inductive step. Intuitively, whenever $\lambda$ is
updated at line $15$, it is as if the unknown target is $U_{\lambda}$ from the
beginning and no inconsistencies arise.

Hence, the learner keeps receiving counterexamples and will never converge to the unknown
target.
\IEEEQED

\begin{algorithm}
\small
\caption{An adversarial teacher in the proof of Theorem
\ref{thm:learning_lpts_undecidable}.}
\label{algo:learning_lpts_undecidable}
\begin{algorithmic}[1]
\STATE $n \leftarrow 1$
\STATE $\lambda \leftarrow$ arbitrary rational in $(0,1)$
\STATE $\ncex \leftarrow \emptyset$, $\pcex \leftarrow \emptyset$
\REPEAT
  \IF {$H_n \not\preorder U_\lambda$}
    \STATE let $N$ be a tree counterexample (Def. \ref{def:cex})
    \STATE $\ncex \leftarrow \ncex \cup \{N\}$
    \STATE return $N$ to the learner as a {\em negative} counterexample
  \ELSIF {$U_\lambda \not\preorder H_n$}
    \STATE let $P$ be a tree counterexample (Def. \ref{def:cex})
    \STATE $\pcex \leftarrow \pcex \cup \{P\}$
    \STATE return $P$ to the learner as a {\em positive} counterexample
  \ELSE
    \STATE $\lambda^+ = \min \left( \{p^\mu_b > \lambda ~|~ \mu \in \text{{\em
Dist}}_a[\ncex]\} \cup \{1\} \right)$
    \STATE $\lambda \leftarrow (\lambda^+ + \lambda)/2$
    \STATE let $P$ be a tree counterexample to $U_\lambda \preorder H_n$ (Def.
\ref{def:cex})
    \STATE $\pcex \leftarrow \pcex \cup \{P\}$
    \STATE return $P$ to the learner as a {\em positive} counterexample
  \ENDIF
  \STATE $n \leftarrow n+1$
\UNTIL {\tt false}
\end{algorithmic}
\end{algorithm}

\subsection{Proof of Theorem \ref{thm:learning_min_lpts_undecidable}}
We give an example where it is impossible for the learner to converge to the
target, up to $\simeq$, in presence of an adversarial teacher.

Consider $H_1$ in Figure
\ref{fig:partition_stochastic} as the unknown target $U$ and let $H_n$ be the hypothesis
at the beginning of each round $n \ge 1$ (we count rounds beginning
with $1$) of the active learning loop. We describe a strategy of a teacher
below to keep generating counterexamples no matter what the conjectured
hypothesis is.

By Condition \ref{cond:learner}, $H_1$ is an LPTS with a single state, which is
also the start state. Initially, in every round $n \ge 1$, the teacher first checks
if $H_n$ has a transition on an action other than $a$, $b$ or $c$ in which case, clearly, $H_n
\not\preorder U$ and a negative tree counterexample is returned using the
algorithm sketched in Section \ref{sec:prelims}. Then, the
teacher checks $P \preorder H_n$ and returns $P$ as a positive tree counterexample if
it fails where $P$ is in Figure \ref{fig:partition_stochastic_cex}.
Note that $P$ has an {\em execution mapping} to $U$ and hence, the
teacher satisfies Condition \ref{cond:teacher_pos}. According to this strategy, the learner
keeps receiving negative counterexamples for transitions on actions other than
$a$, $b$ and $c$ or the positive counterexample $P$ which can go on forever, in
which case we are done, or its hypothesis converges to the LPTS $H^*$
(disallowing duplicate transitions) with a single state and Dirac self-loops on
$a$, $b$ and $c$. We will assume the latter,
\ie the learner conjectures $H^*$ after some finite number of rounds. Note that
it is possible that $P$ has not yet been returned as a positive counterexample
to the learner.

At this point, the teacher returns $N_a$ in Figure
\ref{fig:partition_stochastic_cex} as a negative counterexample. This
forces every future hypothesis to have at least two states. In fact, the
LPTS $H_\lambda$ with two states in Figure
\ref{fig:partition_stochastic}, for any $0 < \lambda < 1$ is a
consistent hypothesis. By Condition \ref{cond:learner}, the next hypothesis has
only two states. Now, we describe the teacher's strategy for future rounds. For
this strategy, we show that a consistent LPTS of two states exists
and that a counterexample can be returned, in every round. So, let $s_1$ and
$s_2$ be the two states of the hypothesis with $s_1$ being the start
state. Furthermore, let $\Delta^i_a$, $\Delta^i_b$, and $\Delta^i_c$ be the sets of distributions outgoing from $s_i$,
$i = 1,2$, on actions $a$, $b$ and $c$, respectively. The teacher's strategy
proceeds in every future round is as follows.

  \begin{enumerate}
  \item As in the initial strategy, it first checks if there is a reachable
state in the hypothesis with a transition on an action other than $a$, $b$ and
$c$ and returns a negative counterexample (see Section \ref{sec:prelims}) if there is one.

  \item Then, it checks $P \preorder H_n$ and returns $P$ as a positive
counterexample if it fails.

  \item At this point, $P \preorder H_n$ and $N_a \not\preorder H_n$ hold
($H_n$ is consistent with them) and we infer the following.

    \begin{enumerate}[(i)]
    \item $\Delta^1_a \neq \emptyset$ and for every $\mu_a \in \Delta^1_a$,
$\mu_a(s_1) < 1$ and
    \item $\Delta^1_b \neq \emptyset$ and for every $\mu_b \in \Delta^1_b$ and
every $s_i \in \supp{\mu_b}$, $\Delta^i_c \neq \emptyset$.
    \end{enumerate}

  The teacher, therefore, does the following.

    \begin{enumerate}
    \item If there is a $\mu_b \in \Delta^1_b$ with $\mu_b(s_1) = 1$, it returns
$N_b$ in Figure \ref{fig:partition_stochastic_cex} as a negative
counterexample. Clearly, $N_b$ has an {\em execution mapping} to $H_n$.
    \item Otherwise, there exists a $\mu_b \in \Delta^1_b$ with $\mu_b(s_2) >
0$, implying $\Delta^2_c \neq \emptyset$ and $N^{\beta,\gamma}_c$ in Figure
\ref{fig:partition_stochastic_cex} is returned as a negative
counterexample, where $\beta = \mu_a(s_2)$ for some $\mu_a \in \Delta^1_a$
and $\gamma = \mu_c(s_2)$ for some $\mu_c \in \Delta^2_c$.
Again, $N^{\beta,\gamma}_c$ has an {\em execution mapping} to $H_n$. 
    \end{enumerate}
  \end{enumerate}

Clearly, except for a counterexample generated in case 3(b) above, $H_\lambda$ is a consistent hypothesis for
any $\lambda \in (0,1)$. For case 3(b), $H_\lambda$ with $0 < \lambda < \beta$ is
consistent. So, after any round, $H_\lambda$ with $\lambda$ set to a value
smaller than the least $\beta$ of any $N^{\beta,\gamma}_c$ returned is consistent
and such a $\lambda$ always exists as there are infinite rationals in $(0,1)$.
Thus, Condition \ref{cond:learner} forces the learner to always conjecture a two
state LPTS and hence, it keeps receiving counterexamples and will never converge
to $U$.
\IEEEQED

%% file: main_arxiv.bbl
\begin{thebibliography}{10}
\providecommand{\url}[1]{#1}
\csname url@samestyle\endcsname
\providecommand{\newblock}{\relax}
\providecommand{\bibinfo}[2]{#2}
\providecommand{\BIBentrySTDinterwordspacing}{\spaceskip=0pt\relax}
\providecommand{\BIBentryALTinterwordstretchfactor}{4}
\providecommand{\BIBentryALTinterwordspacing}{\spaceskip=\fontdimen2\font plus
\BIBentryALTinterwordstretchfactor\fontdimen3\font minus
  \fontdimen4\font\relax}
\providecommand{\BIBforeignlanguage}[2]{{%
\expandafter\ifx\csname l@#1\endcsname\relax
\typeout{** WARNING: IEEEtranS.bst: No hyphenation pattern has been}%
\typeout{** loaded for the language `#1'. Using the pattern for}%
\typeout{** the default language instead.}%
\else
\language=\csname l@#1\endcsname
\fi
#2}}
\providecommand{\BIBdecl}{\relax}
\BIBdecl

\bibitem{AHJ_concur01}
L.~d. Alfaro, T.~A. Henzinger, and R.~Jhala, ``{Compositional Methods for
  Probabilistic Systems},'' in \emph{CONCUR}, ser. LNCS, vol. 2154.\hskip 1em
  plus 0.5em minus 0.4em\relax London, UK: Springer-Verlag, 2001, pp. 351--365.

\bibitem{Angluin_infcomp87}
D.~Angluin, ``{Learning Regular Sets from Queries and Counterexamples},''
  \emph{Information and Computation}, vol. 75(2), pp. 87--106, Nov. 1987.

\bibitem{AS_compsurv83}
D.~Angluin and C.~H. Smith, ``{Inductive Inference: Theory and Methods},''
  \emph{ACM Comp. Surv.}, vol. 15(3), pp. 237--269, September 1983.

\bibitem{BEM_jcss00}
C.~Baier, B.~Engelen, and M.~Majster-Cederbaum, ``{Deciding Bisimilarity and
  Similarity for Probabilistic Processes},'' \emph{J. Comput. Syst. Sci.}, vol.
  60(1), pp. 187--231, Feb 2000.

\bibitem{BBB+_jacm00}
A.~Beimel, F.~Bergadano, N.~H. Bshouty, E.~Kushilevitz, and S.~Varricchio,
  ``{Learning Functions Represented as Multiplicity Automata},'' \emph{J. ACM},
  vol. 47(3), pp. 506--530, May 2000.

\bibitem{CO_rairo99}
R.~C. Carrasco and J.~Oncina, ``{Learning Deterministic Regular Grammars From
  Stochastic Samples in Polynomial Time},'' \emph{RAIRO}, vol.~33, pp. 1--20,
  1999.

\bibitem{COC_ml01}
R.~C. Carrasco, J.~Oncina, and J.~Calera-Rubio, ``{Stochastic Inference of
  Regular Tree Languages},'' \emph{Machine Learning}, vol. 44(1-2), pp.
  185--197, July 2001.

\bibitem{CV_tocl10}
R.~Chadha and M.~Viswanathan, ``{A Counterexample-Guided Abstraction-Refinement
  Framework for Markov Decision Processes},'' \emph{TOCL}, vol. 12(1), pp.
  1--49, November 2010.

\bibitem{CCS+_cav05}
S.~Chaki, E.~M. Clarke, N.~Sinha, and P.~Thati, ``{Automated Assume-Guarantee
  Reasoning for Simulation Conformance},'' in \emph{CAV}, ser. LNCS, vol.
  3576.\hskip 1em plus 0.5em minus 0.4em\relax Springer-Verlag, 2005, pp.
  534--547.

\bibitem{CFC+_tacas09}
Y.-F. Chen, A.~Farzan, E.~M. Clarke, Y.-K. Tsay, and B.-Y. Wang, ``{Learning
  Minimal Separating DFA's for Compositional Verification},'' in \emph{TACAS},
  ser. LNCS, vol. 5505.\hskip 1em plus 0.5em minus 0.4em\relax Berlin,
  Heidelberg: Springer-Verlag, 2009, pp. 31--45.

\bibitem{CLM_lics89}
E.~Clarke, D.~Long, and K.~McMillan, ``{Compositional Model Checking},'' in
  \emph{LICS}.\hskip 1em plus 0.5em minus 0.4em\relax Piscataway, NJ, USA: IEEE
  Press, 1989, pp. 353--362.

\bibitem{CGP_book}
E.~M. Clarke, O.~Grumberg, and D.~A. Peled, \emph{{Model Checking}}.\hskip 1em
  plus 0.5em minus 0.4em\relax Cambridge, MA, USA: MIT Press, 2000.

\bibitem{HO_icgi04}
C.~de~la Higuera and J.~Oncina, ``{Learning Stochastic Finite Automata},'' in
  \emph{ICGI}, ser. LNCS, vol. 3264.\hskip 1em plus 0.5em minus 0.4em\relax
  Springer-Verlag, 2004, pp. 175--186.

\bibitem{FHK+_atva11}
L.~Feng, T.~Han, M.~Kwiatkowska, and D.~Parker, ``{Learning-based Compositional
  Verification for Synchronous Probabilistic Systems},'' in \emph{ATVA}, ser.
  LNCS, vol. 6996.\hskip 1em plus 0.5em minus 0.4em\relax Berlin, Heidelberg:
  Springer-Verlag, 2011, pp. 511--521.

\bibitem{FKP_fase11}
L.~Feng, M.~Kwiatkowska, and D.~Parker, ``{Automated Learning of Probabilistic
  Assumptions for Compositional Reasoning},'' in \emph{FASE}, ser. LNCS, vol.
  6603.\hskip 1em plus 0.5em minus 0.4em\relax Berlin, Heidelberg:
  Springer-Verlag, 2011, pp. 2--17.

\bibitem{GO_report93}
P.~Garcia and J.~Oncina, ``{Inference of Recognizable Tree Sets},'' Universidad
  Politecnica de, Research Report DSIC - II/47/93, 1993.

\bibitem{Gold_infcomp78}
E.~M. Gold, ``{Complexity of Automaton Identification from Given Data},''
  \emph{Information and Control}, vol. 37(3), pp. 302--320, 1978.

\bibitem{GMF_fmsd08}
A.~Gupta, K.~L. McMillan, and Z.~Fu, ``{Automated Assumption Generation for
  Compositional Verification},'' \emph{FMSD}, vol. 32(3), pp. 285--301, June
  2008.

\bibitem{KPC_cav12}
A.~Komuravelli, C.~S. P\u{a}s\u{a}reanu, and E.~M. Clarke, ``{Assume-Guarantee
  Abstraction Refinement for Probabilistic Systems},'' in \emph{CAV}, 2012, (to
  appear).

\bibitem{KNP+_tacas10}
M.~Kwiatkowska, G.~Norman, D.~Parker, and H.~Qu, ``{Assume-Guarantee
  Verification for Probabilistic Systems},'' in \emph{TACAS}, ser. LNCS, vol.
  6015.\hskip 1em plus 0.5em minus 0.4em\relax Berlin, Heidelberg:
  Springer-Verlag, 2010, pp. 23--37.

\bibitem{MCJ+_qest11}
H.~Mao \emph{et~al.}, ``{Learning Probabilistic Automata for Model Checking},''
  in \emph{QEST}.\hskip 1em plus 0.5em minus 0.4em\relax Washington, DC, USA:
  IEEE Computer Society, 2011, pp. 111--120.

\bibitem{Milner_tr71}
R.~Milner, ``{An Algebraic Definition of Simulation between Programs},''
  Stanford, CA, USA, Tech. Rep., 1971.

\bibitem{OM_spire98}
A.~L. Oliveira and J.~P. Marques-Silva, ``{Efficient Search Techniques for the
  Inference of Minimum Size Finite Automata},'' in \emph{SPIRE}.\hskip 1em plus
  0.5em minus 0.4em\relax IEEE Computer Society Press, 1998, pp. 81--89.

\bibitem{OG_asspr92}
J.~Oncina and P.~Garcia, ``{Identifying Regular Languages In Polynomial
  Time},'' in \emph{ASSPR}, vol.~5.\hskip 1em plus 0.5em minus 0.4em\relax
  World Scientific, 1992, pp. 99--108.

\bibitem{Pnueli_lmcs85}
A.~Pnueli, ``{In Transition from Global to Modular Temporal Reasoning about
  Programs},'' in \emph{LMCS}, ser. NATO ASI, vol.~13.\hskip 1em plus 0.5em
  minus 0.4em\relax Springer-Verlag, 1985, pp. 123--144.

\bibitem{PGB+_fmsd08}
C.~S. P\u{a}s\u{a}reanu, D.~Giannakopoulou, M.~G. Bobaru, J.~M. Cobleigh, and
  H.~Barringer, ``{Learning to Divide and Conquer: Applying the L* Algorithm to
  Automate Assume-Guarantee Reasoning},'' \emph{FMSD}, vol. 32(3), pp.
  175--205, June 2008.

\bibitem{Rabin_infctrl63}
M.~O. Rabin, ``{Probabilistic Automata},'' \emph{{Information and Control}},
  vol. 6(3), pp. 230--245, 1963.

\bibitem{SL_nordic95}
R.~Segala and N.~Lynch, ``{Probabilistic Simulations for Probabilistic
  Processes},'' \emph{{Nordic J. of Computing}}, vol. 2(2), pp. 250--273, June
  1995.

\bibitem{Tzeng_ml92}
W.-G. Tzeng, ``{Learning Probabilistic Automata and Markov Chains via
  Queries},'' \emph{Machine Learning}, vol. 8(2), pp. 151--166, March 1992.

\bibitem{Zhang_thesis08}
L.~Zhang, ``{Decision Algorithms for Probabilistic Simulations},'' Ph.D.
  dissertation, Universit\"{ä}t des Saarlandes, 2008.

\end{thebibliography}
